\documentclass[aps,floats]{revtex4}
\usepackage{amsmath,amssymb}
\usepackage{graphicx,epsfig}

\begin{document}
\bibliographystyle {plain}

\def\oppropto{\mathop{\propto}} 
\def\opsimeq{\mathop{\simeq}}
\def\opoverderline{\mathop{\overline}}
\def\operarrow{\mathop{\longrightarrow}}
\def\opsim{\mathop{\sim}}

\def\fig#1#2{\includegraphics[height=#1]{#2}}
\def\figx#1#2{\includegraphics[width=#1]{#2}}


\title{ Anderson localization on the Cayley tree : \\
multifractal statistics of the transmission at criticality and off criticality} 


\author{ C\'ecile Monthus and Thomas Garel }
 \affiliation{Institut de Physique Th\'{e}orique, CNRS and CEA Saclay
91191 Gif-sur-Yvette cedex, France}

\begin{abstract}
In contrast to finite dimensions where disordered systems display multifractal statistics only at criticality, the tree geometry induces multifractal statistics for disordered systems also off criticality. For the Anderson tight-binding localization model defined on a tree of branching ratio $K=2$ with $N$ generations, we consider the Miller-Derrida scattering geometry [J. Stat. Phys. 75, 357 (1994)], where an incoming wire is attached to the root of the tree, and where $K^{N}$ outcoming wires are attached to the leaves of the tree. In terms of the $K^{N}$ transmission amplitudes $t_j$, the total Landauer transmission is $T \equiv \sum_j \vert t_j \vert^2$, so that each channel $j$ is characterized by the weight $w_j=\vert t_j \vert^2/T$. We numerically measure the typical multifractal singularity spectrum $f(\alpha)$ of these weights as a function of the disorder strength $W$ and we obtain the following conclusions for its left-termination point $\alpha_+(W)$. In the delocalized phase $W<W_c$, $\alpha_+(W)$ is strictly positive $\alpha_+(W)>0$ and is associated with a moment index $q_+(W)>1$. At criticality, it vanishes $\alpha_+(W_c)=0$ and is associated with the moment index $q_+(W_c)=1$. In the localized phase $W>W_c$, $\alpha_+(W)=0$ is associated with some moment index $q_+(W)<1$. We discuss the similarities with the exact results concerning the multifractal properties of the Directed Polymer on the Cayley tree. 
\end{abstract}

\maketitle

\section{ Introduction}

Since its discovery fifty years ago \cite{anderson}
Anderson localization has remained a very active field
of research (see for instance the reviews
 \cite{thouless,souillard,bookpastur,Kramer,markos,mirlinrevue}).
According to the scaling theory \cite{scaltheo}, 
there is no delocalized phase in dimensions $d=1,2$,
whereas there exists a localization/delocalization at finite disorder
in dimension $d>2$. To get some insight into this type of transition,
it is natural to consider Anderson localization on the Cayley tree
which is expected to represent some mean-field limit.
The tight-binding Anderson model on the Cayley
 tree has been thus studied by various techniques over the years 
 \cite{abou,Kun_Sou,MirlinBethe,DR,MD,us_cayleytraveling}.
Other studies have focused on random-scattering models on the Cayley tree
\cite{shapiro,chalker,bell}. 
For the version of the model defined on random regular graph of fixed degree,
we refer to the recent work \cite{biroli} and references therein.
The motivation to study Anderson localization on the Cayley tree has
been revived recently by the question of many-body localization
\cite{manybodyloc},  because the geometry of the Fock space of
many-body states was argued to be similar to a Cayley tree
\cite{levitov,silvestrov97,silvestrov98,silvestrov,gornyi,huse,us_manybodyloc}. 
But of course,
the questions on many-body localization are much more difficult and
are still debated in the recent studies
\cite{silvestrov98,huse,us_manybodyloc,prosen,prelovsek,reichman,pal,barisic}. 

In quantum coherent problems, the most appropriate characterisation 
of transport properties consists in defining a scattering problem
where the disordered sample is linked to incoming wires and outgoing wires
and in studying the reflection and transmission coefficients.
This scattering theory definition of transport, 
first introduced by Landauer \cite{landauer},
has been much used for one-dimensional systems \cite{anderson_fisher,anderson_lee,luck}
and has been generalized to higher dimensionalities and multi-probe
measurements (see the review \cite{stone} and references therein).
For the Anderson model on the Cayley tree, an appropriate scattering geometry
has been introduced by Miller and Derrida \cite{MD} to perform 
weak-disorder expansions and numerical computations : 
 an incoming wire is attached to the root, 
and $K^{N}$ outcoming wires are attached to the leaves
of a tree of branching ratio $K$ with $N$ generations.
 In a previous work \cite{us_cayleytraveling}, we have used this scattering geometry
to study numerically the statistical properties of total the Landauer transmission
$T \equiv \sum_j \vert t_j \vert^2$
as a function of the number $N$ of generations and of the disorder strength
and to measure its critical behavior. 
The aim of this paper is to characterize the spatial inhomogeneity
between the various channels $j$ : the weights $\vert t_j \vert^2/T$
of the $K^N$ channels turn out to present a multifractal statistics,
not only at criticality but also in the localized and delocalized phases
as a consequence of the tree geometry. So we analyse 
how the singularity spectrum $f(\alpha)$
changes as a function of the disorder strength.

The paper is organized as follows.
In section \ref{sec_trans}, we introduce the Anderson localization
tight-binding model on the Cayley tree and the scattering geometry that we consider
to study the multifractal statistics of the Landauer transmission.
Our numerical results concerning the multifractal statistics in various phases
are described in sections 
 \ref{sec_box} and \ref{sec_cauchy} for the Box distribution and for the Cauchy distribution of disorder respectively.
Our conclusions are summarized in section \ref{sec_conclusion}.
In Appendix \ref{sec_DP}, we recall the exactly known results concerning
the multifractality for the Directed Polymer on the Cayley tree, 
as a classical model which is useful to consider as a comparison,
both for conceptual and numerical purposes. Appendix \ref{app_numerics} 
explains how the numerical singularity spectra presented on figures
have been obtained.

\section{ Scattering geometry for Anderson localization on the Cayley tree}

\label{sec_trans}

\subsection{Anderson tight-binding model on the Cayley tree}

We consider the Anderson tight-binding model
\begin{eqnarray}
H = \sum_i \epsilon_i \vert i > < i \vert  +  \sum_{<i,j>}  \vert i > < j \vert
\label{handerson}
\end{eqnarray}
where the hopping between nearest neighbors $<i,j>$ is a constant $V=1$ 
and where the on-site energies $\epsilon_i$ are independent random variables
drawn from the 'Box' distribution 
\begin{eqnarray}
p_{Box}(\epsilon) = \frac{1}{W}
 \theta \left( - \frac{W}{2} \leq \epsilon \leq \frac{W}{2}  \right)
\label{box}
\end{eqnarray}
The parameter $W$ thus represents the disorder strength.
We have also studied the case of the Cauchy disorder
\begin{eqnarray}
p_{Cauchy}(\epsilon) = \frac{ W }{ \pi (\epsilon^2 + W^2) }
\label{cauchy}
\end{eqnarray}

\subsection{Miller-Derrida scattering geometry }

\begin{figure}[htbp]
 \includegraphics[height=6cm]{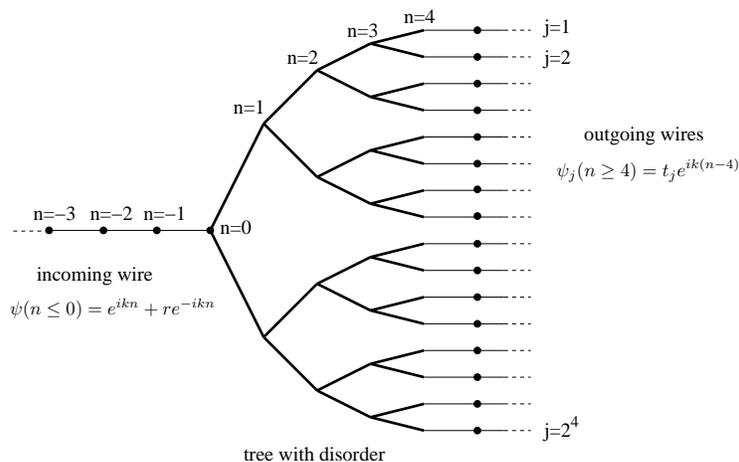}
\caption{ Scattering geometry of Ref. \cite{MD} : the disordered tree 
of branching ratio $K=2$ starting at generation $n=0$
 and ending at generation $N$ (on the Figure $N=4$)
is attached to one incoming wire and to $K^{N}$ outgoing wires. 
The total transmission is 
$T \equiv \sum_j \vert t_j \vert^2 = 1 - \vert r \vert^2 $
where $r$ is the reflection amplitude of the incoming wire, 
and $t_j$ the transmission
amplitudes of the outgoing wires.}
\label{figscattering}
\end{figure}

We consider the scattering geometry introduced in \cite{MD}
and shown on Fig. \ref{figscattering} :
 the finite tree of branching ratio $K$ 
is attached to one incoming wire at its root (generation $n=0$)
and to $K^{N}$ outgoing wires at generation $N$.
One is interested into the eigenstate $\vert \psi >$ that satisfies 
the  Schr\"odinger equation
\begin{eqnarray}
H \vert \psi > = E  \vert \psi > 
\label{schodinger}
\end{eqnarray}
inside the disorder sample and in the wires
 where one requires the plane-wave forms
\begin{eqnarray}
\psi(n \leq 0) && = e^{ik n} +r e^{- i k n} \nonumber \\
 \psi_j(n \geq  N) && = t_j e^{ik (n- N)} 
\label{psiqoutside}
\end{eqnarray}
These boundary conditions define
 the reflection amplitude $r$ of the incoming wire
and the transmission amplitudes $t_j$ of the $j=1,2,..K^{N}$ outgoing wires.
To satisfy the Schr\"odinger 
Equation of Eq. \ref{schodinger} within the wires with
the forms of Eq. \ref{psiqoutside}, one has the following relation between
the energy $E$ and the wave vector $k$  
\begin{eqnarray}
 E=2 \cos k  
\label{relationEk}
\end{eqnarray}
To simplify the discussion, we will focus in this paper on the case of
zero-energy $E=0$ and wave-vector $k=\pi/2$,
because the zero-energy $E=0$ corresponds to the center of the band 
where the delocalization first appears when the strength $W$ of the disorder
is decreased from the strong disorder localized phase.
From the conservation of energy, the total transmission $T$ is related to the 
reflection coefficient $\vert r \vert^2$
\begin{eqnarray}
T \equiv \sum_j \vert t_j \vert^2 = 1 - \vert r \vert^2
\label{deftotaltrans}
\end{eqnarray}

We refer to \cite{MD} for the results of
a weak disorder expansion within this framework,
and for a numerical Monte-Carlo approach to determine the mobility edge
in the plane $(E,W)$.
In \cite{us_cayleytraveling} we have studied the statistical properties
over the disordered samples
of the total Landauer transmission $T_N$ 
at zero energy $E=0$ as a function of the disorder
strength $W$ and of the number $N$ of generations. 
In the localized phase $W>W_c$, the typical transmission $T_N^{typ} \equiv e^{\overline{ \ln T_N}}$ decays exponentially  with the number $N$ of generations
\begin{eqnarray}
\ln (T_N^{typ}) \equiv \overline{ \ln T_N(W>W_c) } \opsimeq_{N \to \infty} - \frac{N}{\xi_{loc}(W)}
\label{transloc}
\end{eqnarray}
where $\xi_{loc}$ represents the localization length. 
In the delocalized phase, the typical transmission remains finite 
in the limit where the number of generations $N$ diverges  
\begin{eqnarray}
T_N^{typ} \equiv e^{\overline{ \ln T_N(W<W_c,N) }} \opsimeq_{N \to \infty} 
  T_{\infty}(W<W_c)  > 0
\label{transdeloc}
\end{eqnarray}
The total Landauer transmission $T$ is thus an appropriate
 order parameter of the localization transition at the mobility edge $W_c$.
We refer to \cite{us_cayleytraveling} for more details on the critical behaviors of the localization length $\xi_{loc}$ and of the asymptotic value $ T_{\infty}(W<W_c)$. In the present paper, we wish to analyse the statistics of the 
contributions $t_j$ of the various channels to the total transmission
of Eq. \ref{deftotaltrans} as we now explain.

\subsection{ Statistical properties of the weights of the outgoing channels }

In each disordered sample, we consider the $K^N$ weights
\begin{eqnarray}
w_j \equiv \frac{ \vert t_j \vert ^2}{ T } 
= \frac{ \vert t_j \vert ^2}{\sum_{j'} \vert t_{j'} \vert^2} 
\label{wjtrans}
\end{eqnarray}
and the 'analogs' of Inverse Participation Ratios
\begin{eqnarray}
I_q(M=K^N) \equiv \sum_{j=1}^{M}  w_j^q 
= \frac{\sum_{j=1}^{M} \vert t_j \vert ^{2q} }{\left( \sum_{j=1}^{M} \vert t_j \vert ^{2}\right)^q}
\label{iqdef}
\end{eqnarray}
It is useful to introduce the multifractal formalism 
with respect to $M=K^N$ (or equivalently the large deviation formalism
with respect to the variable $N= (\ln M)/(\ln K)$) :
one defines the typical exponents  $ \tau^{typ}(q) $ as the exponents governing
the decays of the typical values
\begin{eqnarray}
I_q^{typ} (M=K^N)  \equiv e^{\overline{ \ln I_q (M)} } 
 \oppropto_{M \to +\infty} M^{- \tau^{typ}(q)} = e^{- N (\ln K) \tau^{typ}(q)}
\label{tauq}
\end{eqnarray}
The typical singularity spectrum $f^{typ}(\alpha)$ is defined as follows :
in a large disordered sample, the number ${\cal N}_M(\alpha)$ of channels $j$ 
(among the total of $M$ of channels) that have 
a weight $w_j$ scaling as $w_j \sim M^{-\alpha}$ scales as 
\begin{eqnarray}
{\cal N}_M^{typ}(\alpha) \simeq M^{f^{typ}(\alpha)}
\label{calnalpha}
\end{eqnarray}
Then saddle-point computation of $I_q$ (Eqs \ref{iqdef} and \ref{tauq})
\begin{eqnarray}
I_q^{typ}(M=K^N) = \sum_{j=1}^{M} w_j^q \sim \int d\alpha M^{f^{typ}(\alpha) - q \alpha}
\label{saddle}
\end{eqnarray}
leads to the Legendre transform formula
\begin{eqnarray}
- \tau^{typ}(q) = {\rm max}_{\alpha} \left[ f^{typ}(\alpha) - q \alpha \right]
\label{legendre}
\end{eqnarray}

Let us now briefly recall some basic notions about multifractality
that will be useful to analyse the numerical results.
As a consequence of the weight definition of Eq. \ref{wjtrans}, the index $\alpha$ 
cannot be negative, so one has $\alpha \geq 0$.
As a consequence of Eq. \ref{calnalpha}, 
the 'typical' singularity spectrum is non-negative : $f^{typ}(\alpha) \geq 0 $.
[Note that this is in contrast with the 'averaged' singularity spectrum $f^{av}(\alpha) $
which can become negative to describe rare events (see \cite{mirlinrevue} for more details),
but in this paper we only consider the typical singularity spectrum].
The terminations points $\alpha_{\pm}$ are defined as the points where
the singularity spectrum vanishes $f(\alpha_{\pm})=0$, whereas the singularity spectrum 
remains strictly positive in between 
\begin{eqnarray}
 f^{typ}(\alpha) >0 \ \ {\rm for } \ \ \alpha_+ < \alpha < \alpha_-
\label{termination}
\end{eqnarray}
The left termination point $\alpha_+$ which represents the smallest possible $\alpha$
will play an essential role in the following.
From the point of view of the Legendre transform formula of Eq. \ref{legendre},
it is associated with some positive value $q_+>0$,
 where the saddle point $\alpha(q)$ reaches $\alpha_+$,
so that for all higher $q$, the saddle point remains frozen at this value  
\begin{eqnarray}
 \alpha(q>q_+) = \alpha_+ 
\label{alphaqplus}
\end{eqnarray}
 and the typical exponent $\tau^{typ}(q)$ is simply
\begin{eqnarray}
 \tau^{typ}(q>q_+) = q \alpha_+ 
\label{tautypqplus}
\end{eqnarray}
The same discussion can be transposed to the right termination point $\alpha_-$ associated with some negative index $q_-<0$, with $\alpha(q<q_-) = \alpha_- $ and $\tau^{typ}(q<q_-) = q \alpha_- $.
The value $q=0$ is associated with 
the most probable value $\alpha_0 \equiv \alpha(q=0)$
where the singularity spectrum reaches its maximum 
\begin{eqnarray}
f(\alpha_0)=1
\label{analphaq0}
\end{eqnarray}
Finally the value $q=1$ is associated with the value  
$\alpha_1 \equiv \alpha(q=1)$
where the singularity spectrum satisfies 
\begin{eqnarray}
f(\alpha_1)=\alpha_1
\label{tanalphaq1}
\end{eqnarray}
as a consequence of the normalization $I_{q=1}^{typ} = 1 $ corresponding to $\tau^{typ}(q=1)=0$.

\subsection{ Comparison with Anderson localization models in finite dimension }

We should stress here the similarities and differences 
with the usual multifractal definitions used for Anderson localization
models in finite dimension $d$ (see the review \cite{mirlinrevue}) :
for a normalized eigenfunction on the volume $V=L^d$
(i.e. $\sum_{r \in V=L^d} \vert \psi(r) \vert^{2} =1 $), the Inverse Participation
ratios are defined as
\begin{eqnarray}
P_q \equiv \sum_{r \in V=L^d} \vert \psi(r) \vert^{2q}
\label{pqloc}
\end{eqnarray}
and the exponents $\tau(q)$ are defined as 
\begin{eqnarray}
P_q^{typ}  \oppropto_{L \to +\infty} L^{- \tau^{typ}(q)}
\label{finitedloc}
\end{eqnarray}
In finite dimension $d$, powers of $L$ and powers of the volume $V=L^d$
correspond to the same scaling (up to a redefinition of the exponents), 
while on the tree, one should use powers of the 
number $M=K^N$ in the definitions of Eq. \ref{tauq},
and not powers of the linear distance $N$.

Another difficulty with the tree geometry is that sites of different generations
are not equivalent in the pure case (see \cite{us_cayleytraveling} for
explicit expressions of wavefunctions that decay exponentially with the 
distance $N$ in the pure case). This shows that direct generalizations
of $P_q$ where the sum is over all sites of the tree is not appropriate
(see again \cite{us_cayleytraveling} for more detailed discussion on
the anomalous behavior of usual I.P.R.s), and this is why we have chosen 
to consider the weights of the channels in the Miller-Derrida geometry,
since they involve the wave-function weights of the $K^N$ points
that are at the same distance $N$ of the origin.
In the pure case, these weights have all the same weights 
 $\vert t_j \vert^2/T=1/K^N$ (see again \cite{us_cayleytraveling} for more details), leading to the mono-fractal behavior of the $I_q$ of Eq. \ref{iqdef} 
\begin{eqnarray}
I_q^{pure} (M=K^N) 
= \frac{\sum_{j=1}^{K^N} \vert t_j \vert ^{2q} }
{\left( \sum_{j=1}^{K^N} \vert t_j \vert ^{2}\right)^q} = \frac{1}{K^{N(q-1)}}
\label{iqpure}
\end{eqnarray}
So the tree geometry has the peculiarity to induce multifractal behavior
of the $I_q$ even in the delocalized phase (the radial symmetry of the pure case
is not able to survive even at small disorder), whereas in any finite dimension,
the I.P.R. in the delocalized phase are monofractal with the same scaling as the pure case.

Finally in finite dimension, the localized phase is characterized by 
localized eigenfunctions where some rare sites have finite weights,
whereas most sites have exponentially-small weights in the linear size $L$.
On the tree, all eigenfunctions have to decay exponentially with the distance $N$,
 even in the pure case, to fullfill the normalization constraint with an exponentially-growing number of sites with the distance $N$.  
So in the localized phase, the fluctuations of the weights will be also
characterized by a multifractal statistics of the $I_q$ of Eq. \ref{iqdef}.

In summary, in contrast to finite dimensions
 where disordered systems display multifractal statistics only at criticality, 
the tree geometry induces multifractal statistics
for disordered systems also outside criticality, if one consider 
the inhomogeneities among the points at a given distance from the center.
 In the recent mathematical study \cite{aizenman}, similar observables 
have been introduced with a large deviation analysis in $N$.

Since these multifractal properties of disordered models defined on trees
are unusual with respect to finite dimensions, it is useful
to see how the multifractal analysis of Eq. \ref{tauq} works in
an exactly solved model : in the Appendix \ref{sec_DP}
we thus recall the case of the Directed Polymer 
on the Cayley tree, which is a classical disordered model 
having the same geometry.

\section{ Numerical results for the Box distribution }

\label{sec_box}

In this section, we describe our numerical results for a tree of branching ratio $K=2$
where the disordered on-site energies are drawn
with the Box distribution of Eq. \ref{box}.
The critical disorder width $W_c$ at the center of the band $E=0$ has been
found to be numerically in the interval \cite{abou,us_cayleytraveling,biroli}
\begin{eqnarray}
16 < W_c < 18
\label{wcboxinter} 
\end{eqnarray}

\subsection{ Numerical details }

\label{numerical}

We have studied trees 
containing $N$ generations with a corresponding number $n_s(N)$
of disordered samples with the values
\begin{eqnarray}
N && = 10 ; 12 ; 14 ; 16 ; 18; 20; 22; 24 \nonumber \\
n_s(N) && = 10^7 ; 27.10^5 ; 7.10^5 ; 17.10^4 ; 43.10^3 ; 10^4 ; 27.10^2 ; 650
\label{numerics} 
\end{eqnarray}
We have chosen to work only with even $N$, because in the pure case, the total Landauer transmission is perfect $T^{pure}_N=1$ only for even N
(see section 2.2 of Ref \cite{us_cayleytraveling} for more details).
The transition amplitudes $t_j$ of the scattering eigenvalue problem of Eqs
\ref{schodinger} and \ref{psiqoutside} are computed via the introduction of Riccati
variables as explained in details in \cite{MD,us_cayleytraveling}.
The multifractal spectrum is then obtained via the standard method of Ref \cite{chh},
where the curve $f(\alpha)$ is obtained parametrically in the parameter $q$
(see Appendix B \ref{app_numerics} for more details) :
here we have used values in the range $-5 \leq q \leq +5$. 
As shown in Appendix \ref{sec_DP}, 
we have checked that the sizes and statistics of Eq. \ref{numerics} 
were  sufficient to obtain reliable results
 for the multifractal properties of the 
Directed Polymer model by a direct comparison
 with exactly known results in various phases.

Let us make some final remark to explain the differences 
with respect to the numerical method used in our previous work 
concerning the full transmission $T$.
In \cite{us_cayleytraveling}  
we had used the so called 'pool method' (see section 2.3.1 of \cite{us_cayleytraveling})
which allows to study much bigger sizes (like $N \sim 10^5$ generations).
This was possible because the full transmission $T$ can be directly computed from the reflection coefficient of the incoming wire alone
(see Eq. (17) of \cite{us_cayleytraveling}),
i.e. one does not need to compute the whole set of
transmissions $t_j$ of individual channels.
So the full transmission $T$ can be obtained directly from the stable probability distribution of the complex Riccati variables, for which the pool method is well adapted.
However the multifractal spectrum is a much more complicated observable :
it does not depend only on the one-point distribution of the Riccati variables, 
but it involves the whole correlations between the Riccati variables
along branches (each Riccati variable is computed from its $K$ descendants, see Eq. (25)
of \cite{us_cayleytraveling}). Of course one could try to develop some generalized pool
method to compute multifractal properties, but one should be very careful to avoid
artifacts. In the present paper, we have thus chosen to work only
 with exact numerical data on finite trees to avoid any doubts on 
the numerical results.

\subsection{ Delocalized phase }

\begin{figure}[htbp]
 \includegraphics[height=6cm]{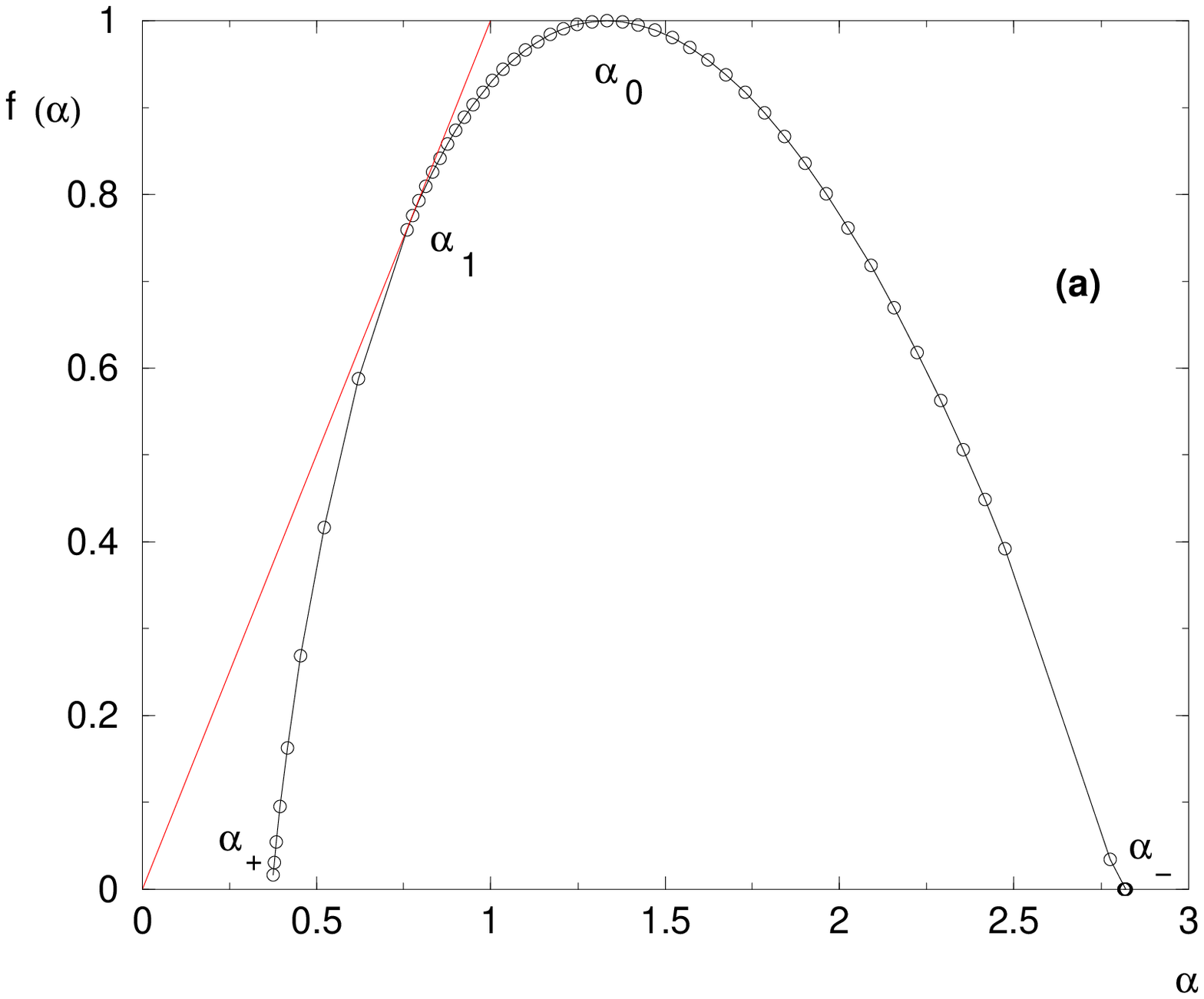}
\hspace{1cm}
 \includegraphics[height=6cm]{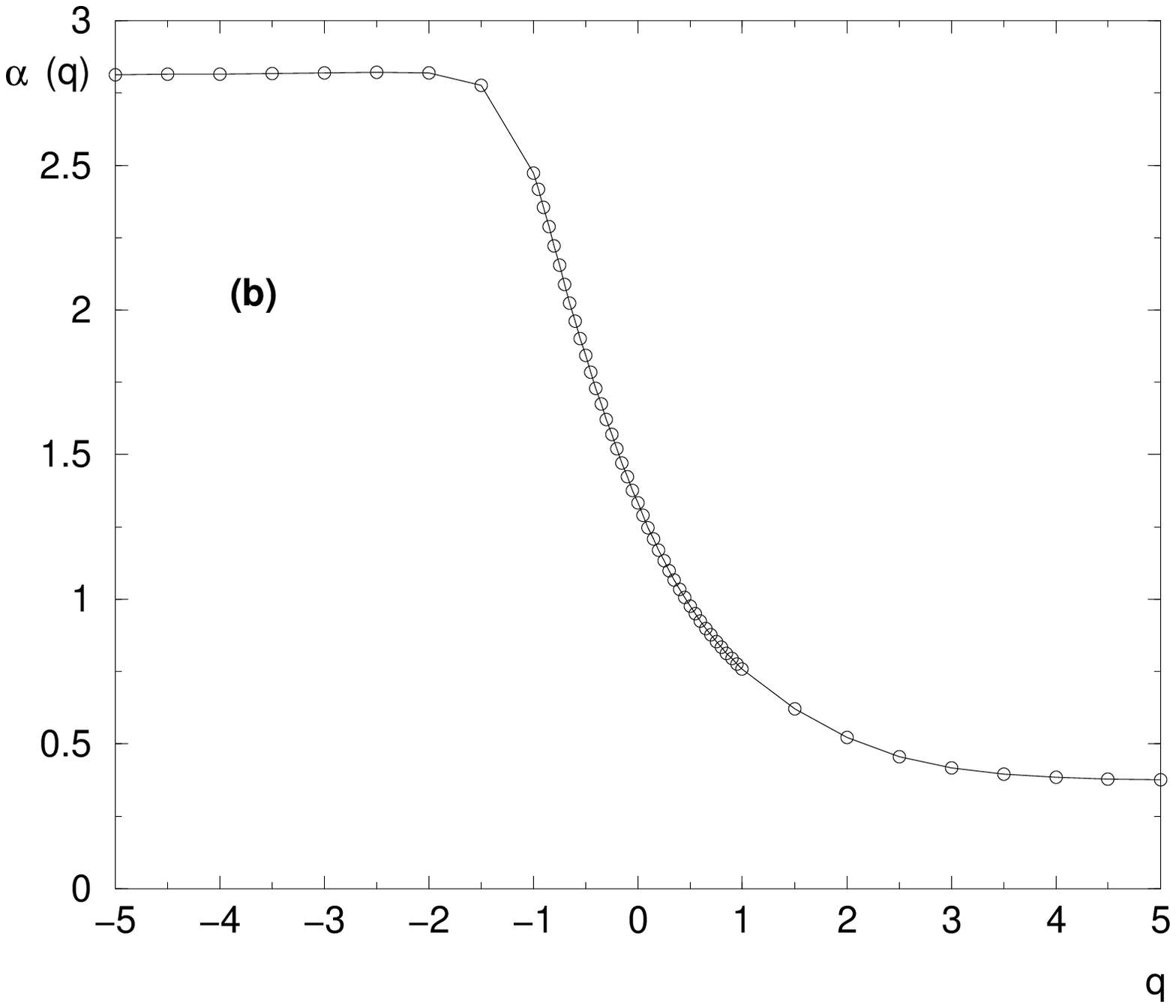}
\caption{ Box disorder $W=5$
(a) The singularity spectrum  $f(\alpha)$ has for termination points $\alpha_+ \simeq 0.37$ and $\alpha_- \simeq 2.82$,
for typical value $\alpha_0 \simeq 1.33$ and for tangent point $\alpha_1=f(\alpha_1) \simeq 0.76$
(b) The corresponding  $\alpha(q)$ saturates at a value around $q_+ \simeq 3$.
  }
\label{figboxwd=5}
\vspace{1cm} 
 \includegraphics[height=6cm]{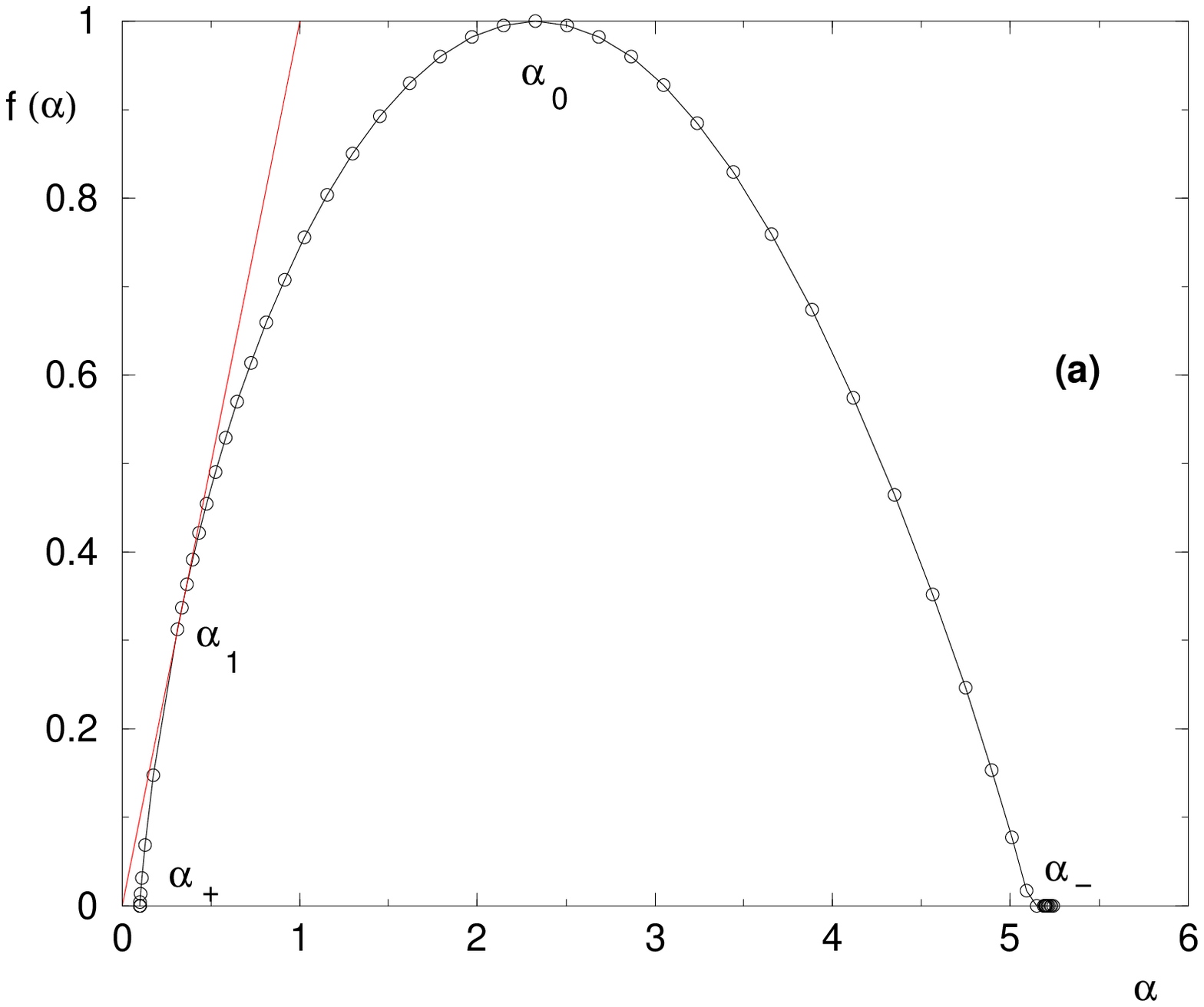}
\hspace{1cm}
 \includegraphics[height=6cm]{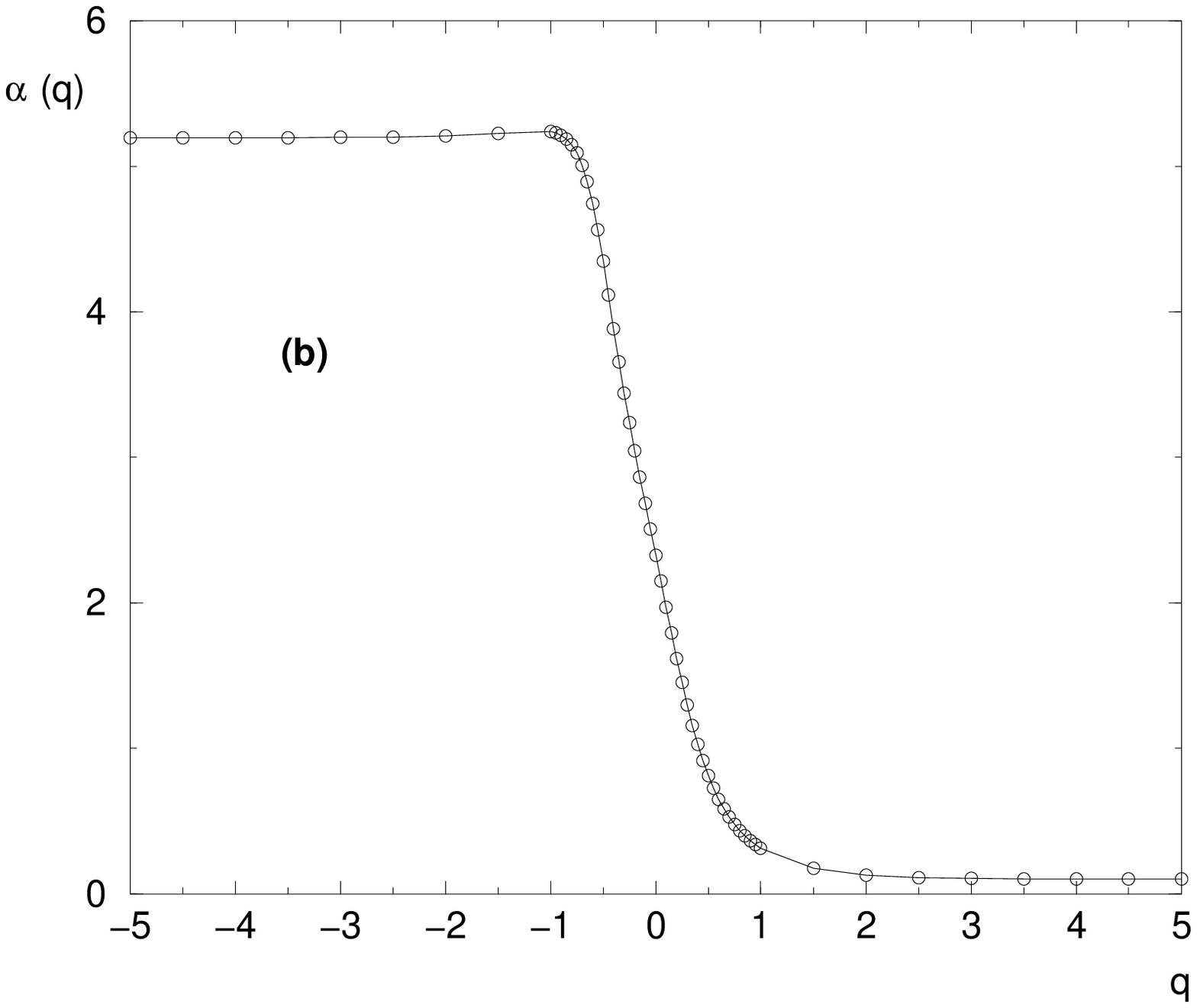}
\caption{ Box disorder $W=10$
(a) The singularity spectrum  $f(\alpha)$ has for termination points $\alpha_+ \simeq 0.1$ and $\alpha_- \simeq 5.2$,
for typical value $\alpha_0 \simeq 2.33$ and for tangent point $\alpha_1=f(\alpha_1) \simeq 0.31 $
(b) The corresponding  $\alpha(q)$ saturates at a value around $q_+ \simeq 2$.
  }
\label{figboxwd=10}
\end{figure}

In the delocalized phase $W<W_c$, we find that the left termination point is strictly positive
$\alpha_+(W)>0$ and is associated with a moment index $q_+(W)>1$.
Two examples of our numerical data are shown on Fig. \ref{figboxwd=5} and \ref{figboxwd=10}
corresponding to $W=5$ and $W=10$ respectively.

\subsection{ Critical point }

\begin{figure}[htbp]
 \includegraphics[height=6cm]{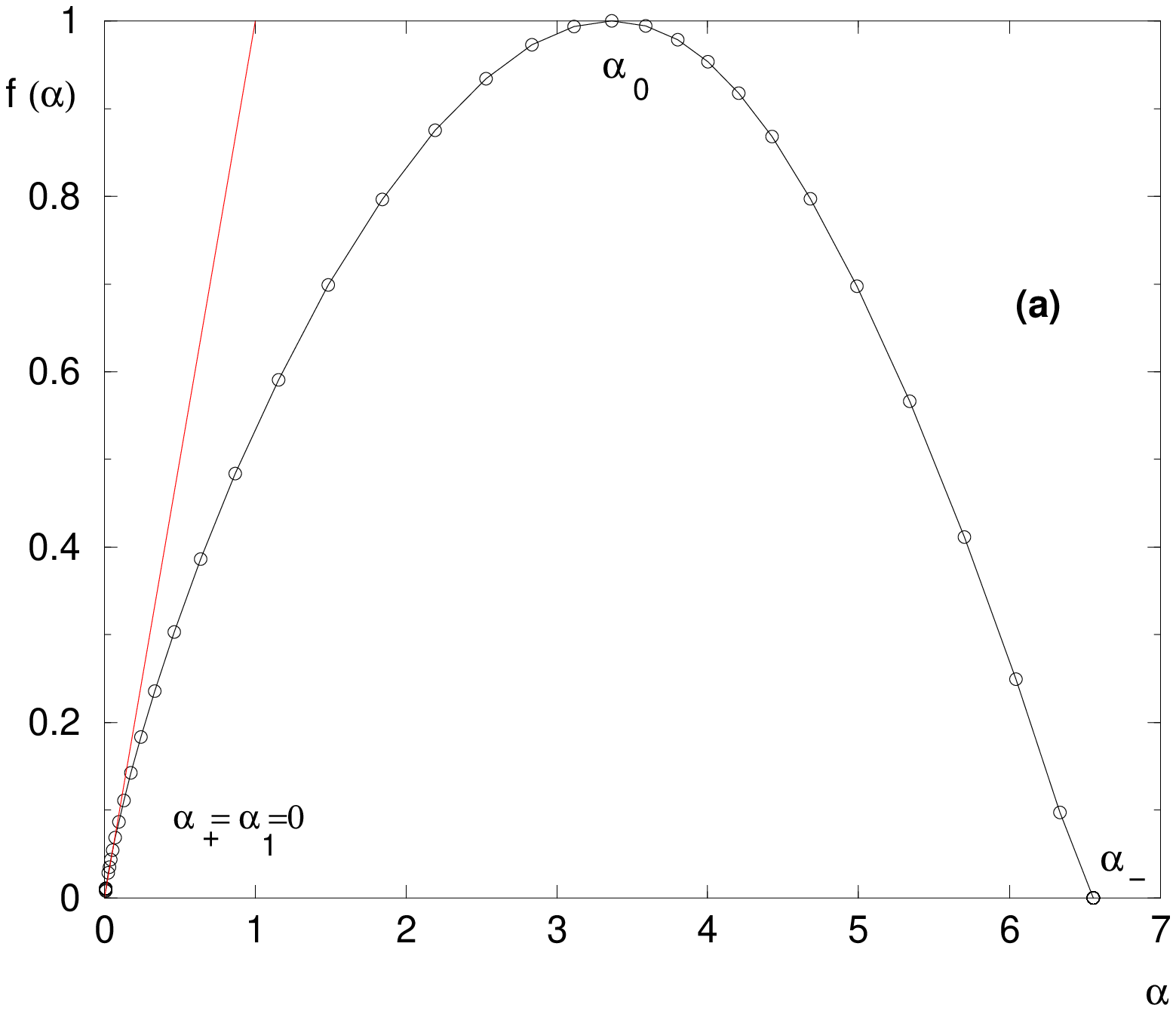}
\hspace{1cm}
 \includegraphics[height=6cm]{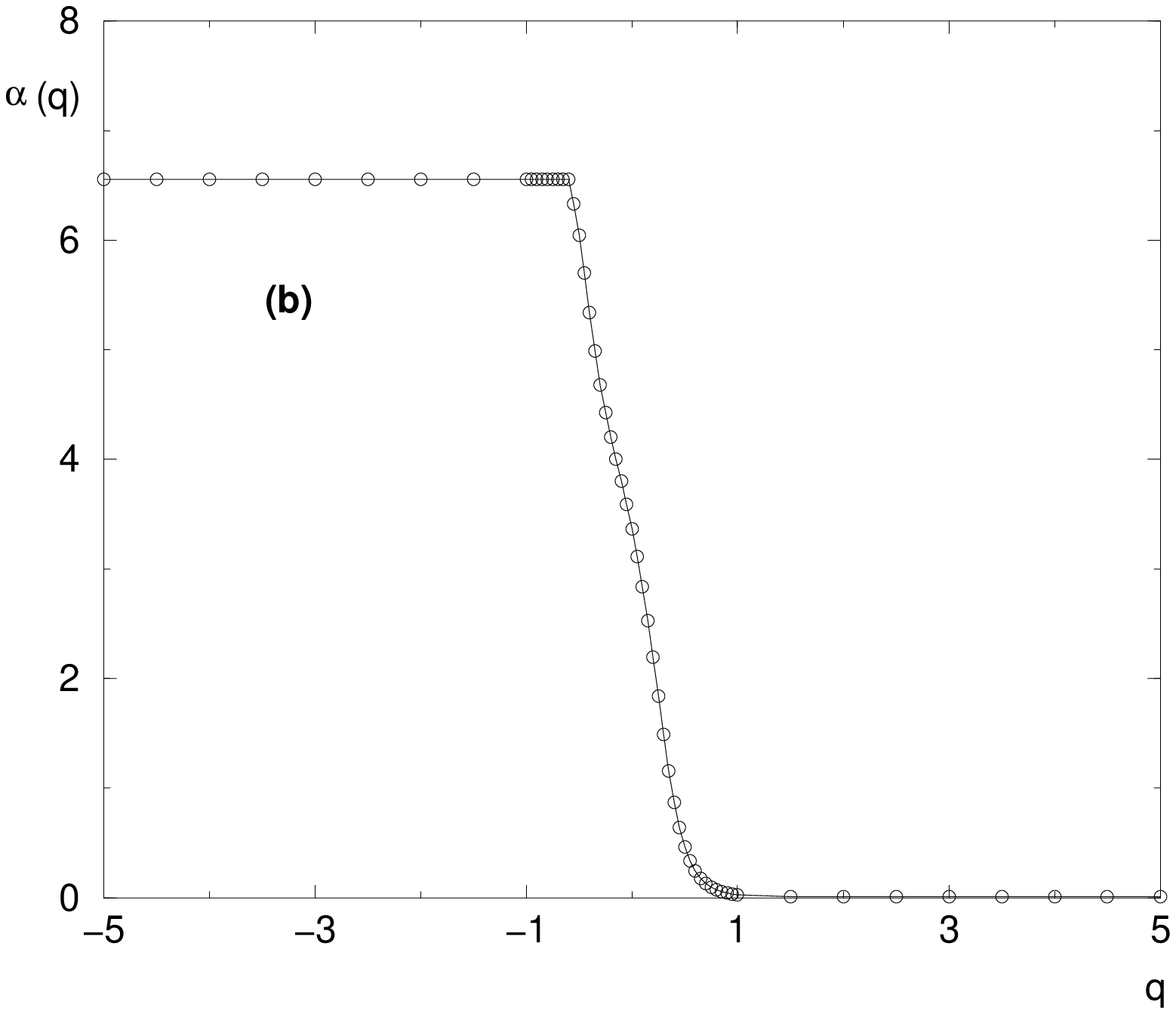}
\caption{ Box disorder $W=17$
(a) The singularity spectrum  $f(\alpha)$ has for termination points $\alpha_+ \simeq 0$ and $\alpha_- \simeq 6.7$,
for typical value $\alpha_0 \simeq 3.36$ and for tangent point $\alpha_1=f(\alpha_1) \simeq 0$
(b) The corresponding  $\alpha(q)$ saturates at the value $q_+ \simeq 1$.
  }
\label{figboxwd=17}
\end{figure}

At criticality, the left termination point vanishes $\alpha_+(W_c)=0$
together with the tangent point $\alpha_1=f(\alpha_1) = 0$, as shown on Fig. \ref{figboxwd=17}
corresponding to $W=17$. The corresponding saddle-point $\alpha(q)$ saturates at the value
$q_+(W_c) \simeq 1$.

\subsection{ Localized phase }

\begin{figure}[htbp]
 \includegraphics[height=6cm]{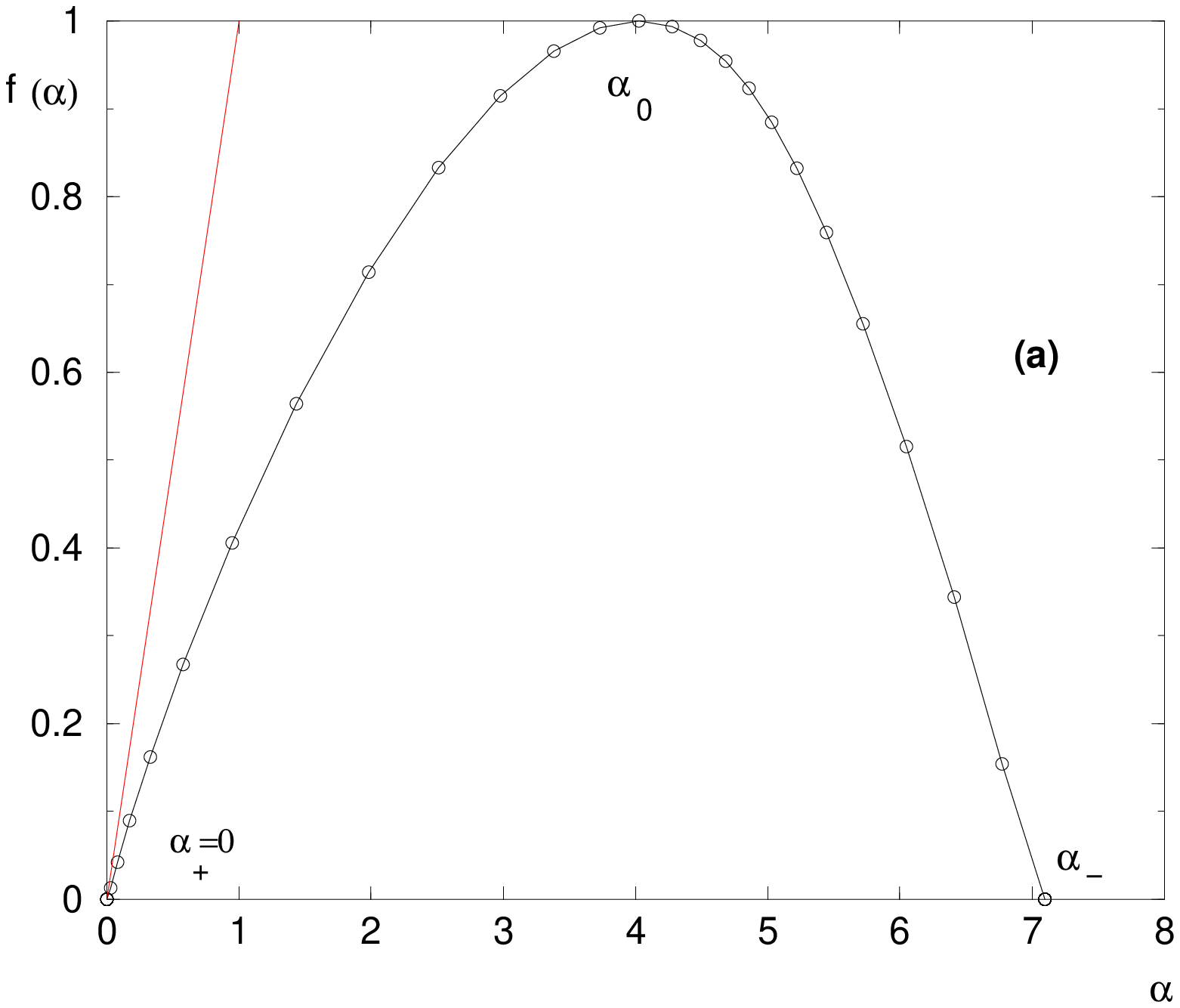}
\hspace{1cm}
 \includegraphics[height=6cm]{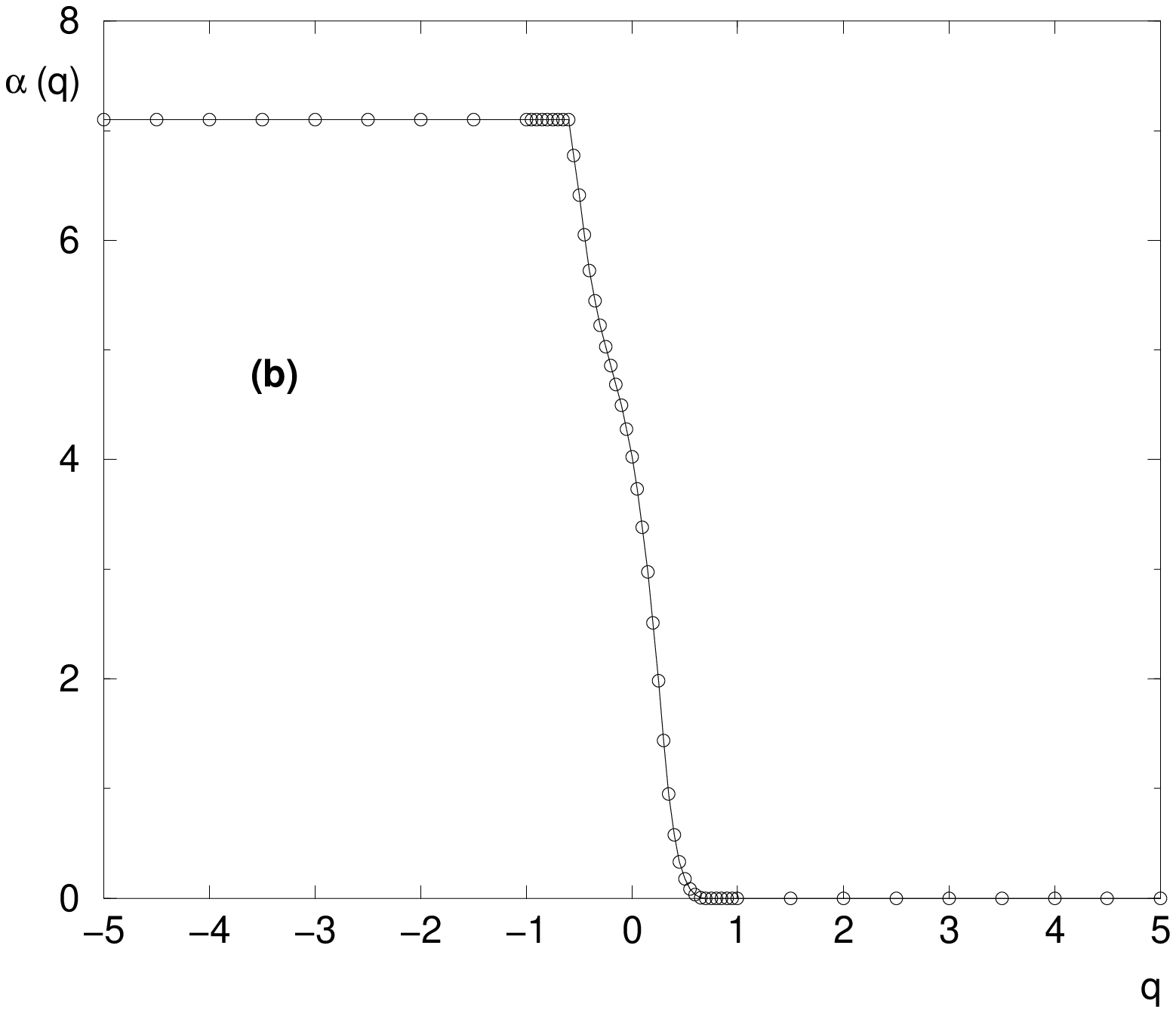}
\caption{ Box disorder $W=30$
(a) The singularity spectrum  $f(\alpha)$ has for termination points $\alpha_+ = 0$ and $\alpha_- \simeq 7.1$,
and for typical value $\alpha_0 \simeq 4.0$
(b) The corresponding  $\alpha(q)$  saturates at the value $q_+ \simeq 0.55$.
  }
\vspace{1cm} 
\label{figboxwd=30}
 \includegraphics[height=6cm]{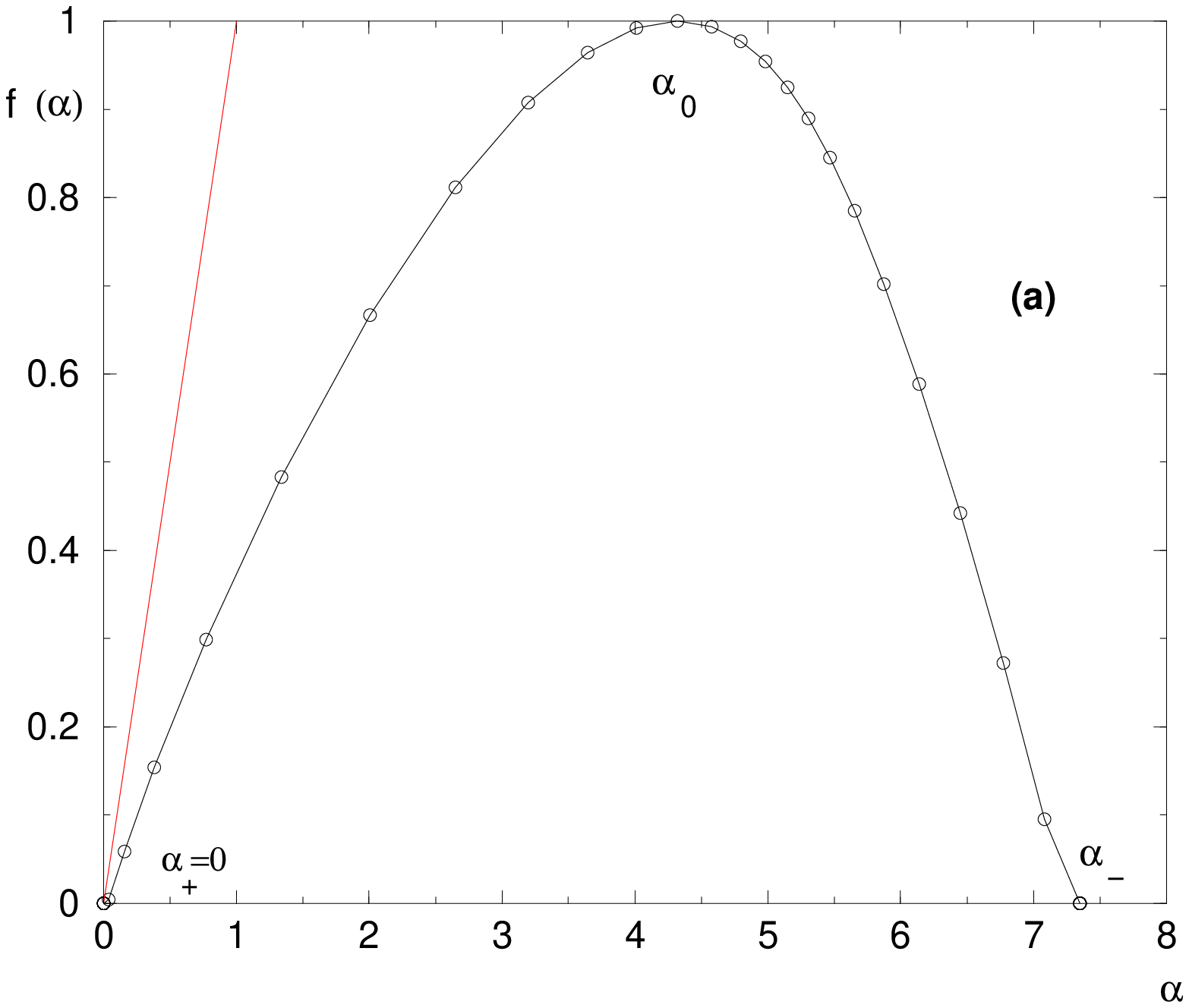}
\hspace{1cm}
 \includegraphics[height=6cm]{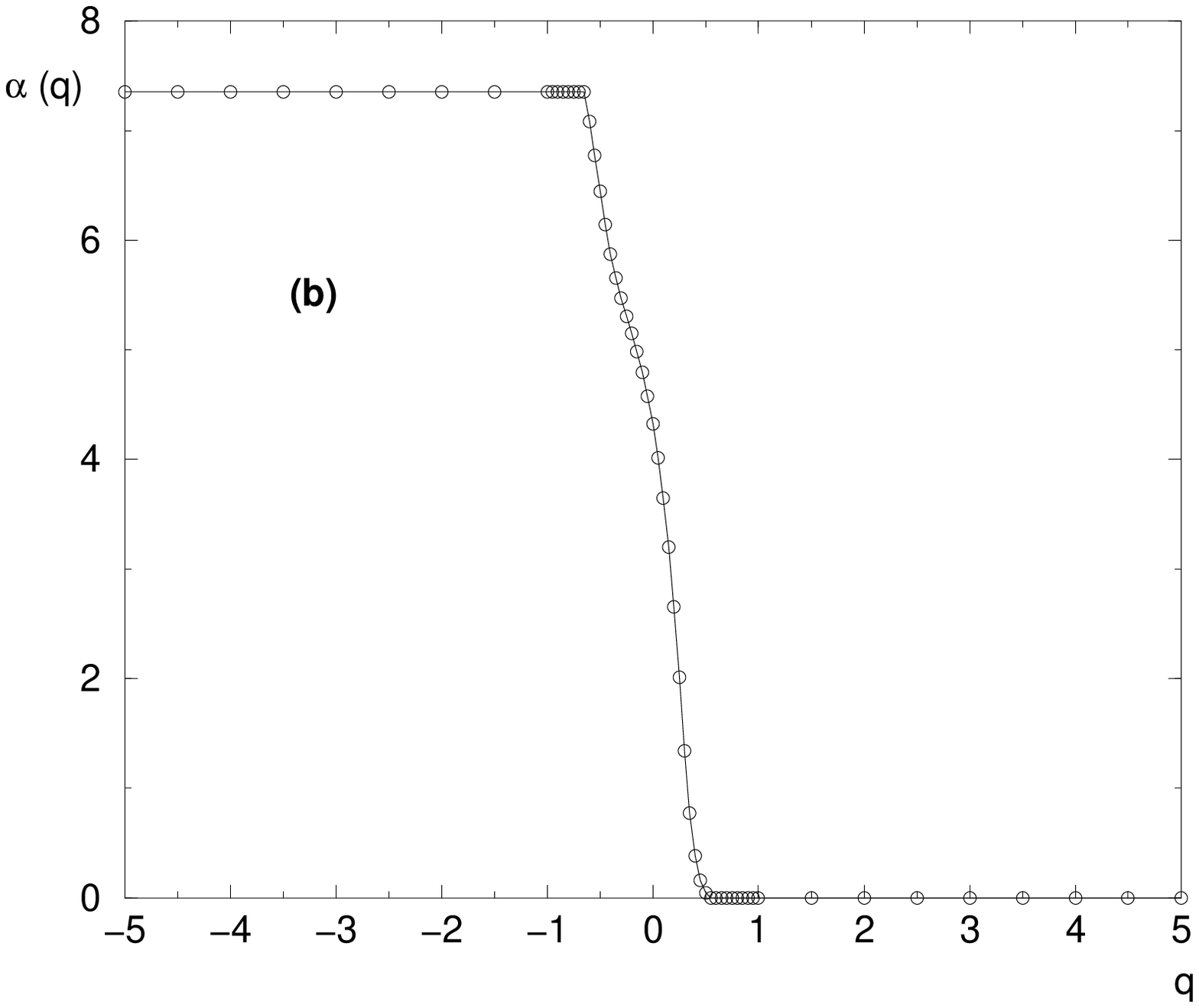}
\caption{ Box disorder $W=40$
(a) The singularity spectrum  $f(\alpha)$ has for termination points $\alpha_+ \simeq 0$ and $\alpha_- \simeq 7.35$,
and for typical value $\alpha_0 \simeq 4.3 $.
(b) The corresponding  $\alpha(q)$  saturates at the value $q_+ \simeq 0.45$.
  }
\label{figboxwd=40}
\end{figure}

 In the localized phase $W>W_c$, the 
vanishing left termination point $\alpha_+(W_c)=0$ is associated
with some moment index $q_+(W)<1$, as shown on Fig. \ref{figboxwd=30} and \ref{figboxwd=40}
corresponding to $W=30$ and  $W=40$ respectively.

 \newpage

\section{ Numerical results for the Cauchy distribution }

\label{sec_cauchy}

In this section, we describe our numerical results for a tree of branching ratio $K=2$
where the disordered on-site energies are drawn
with the Cauchy distribution of Eq. \ref{cauchy}. 
The numerical details are the same as in the section \ref{numerical}.
The critical disorder width $W_c$ at the center of the band $E=0$ has been
previously found to be numerically in the interval \cite{abou}
\begin{eqnarray}
3.6 < W_c < 4.4
\label{wccauchyinter} 
\end{eqnarray}
In many areas, the Cauchy distribution whose variance is infinite
leads to anomalous results with respect to bounded distributions. 
In the context of Anderson localization, the Cauchy disorder
 is of course anomalous from the point of view of weak-disorder expansion 
which contains explicitly the variance of the disorder (see \cite{luck} and references therein). However from the point of view of Anderson localization
transitions at finite disorder, we are not aware of any statement concerning 
its anomalous behaviors (except old conclusions concerning the absence of transition that have been shown to be false afterwards). On the theoretical side, the Cauchy distribution is well known to have many advantages : it is the only 
disorder distribution that leads to an exact and simple solution
in one dimension (see \cite{luck} and references therein), 
and that leads to an exact and simple solution for the density of states in any dimension \cite{llyod}. For the 
Anderson localization on the Caylee tree
 that we consider in the present paper, it is also the only 
disorder distribution that leads to an exact and simple solution
 for the stationary distribution of the Riccati variables
 in the localized phase \cite{abou,MD}, the only remaining problem being
 that it is not known in the delocalized phase where
the Riccati variables are complex \cite{MD}.
These theoretical advantages of the Cauchy distribution justify  
to study numerically its properties and to compare with the case of bounded
distributions. In the following, we obtain that the results concerning the multifractal properties of the transmission for the Cauchy distribution are qualitatively the same as the results obtained in the previous section concerning the box distribution.

\subsection{ Delocalized phase }

\begin{figure}[htbp]
 \includegraphics[height=6cm]{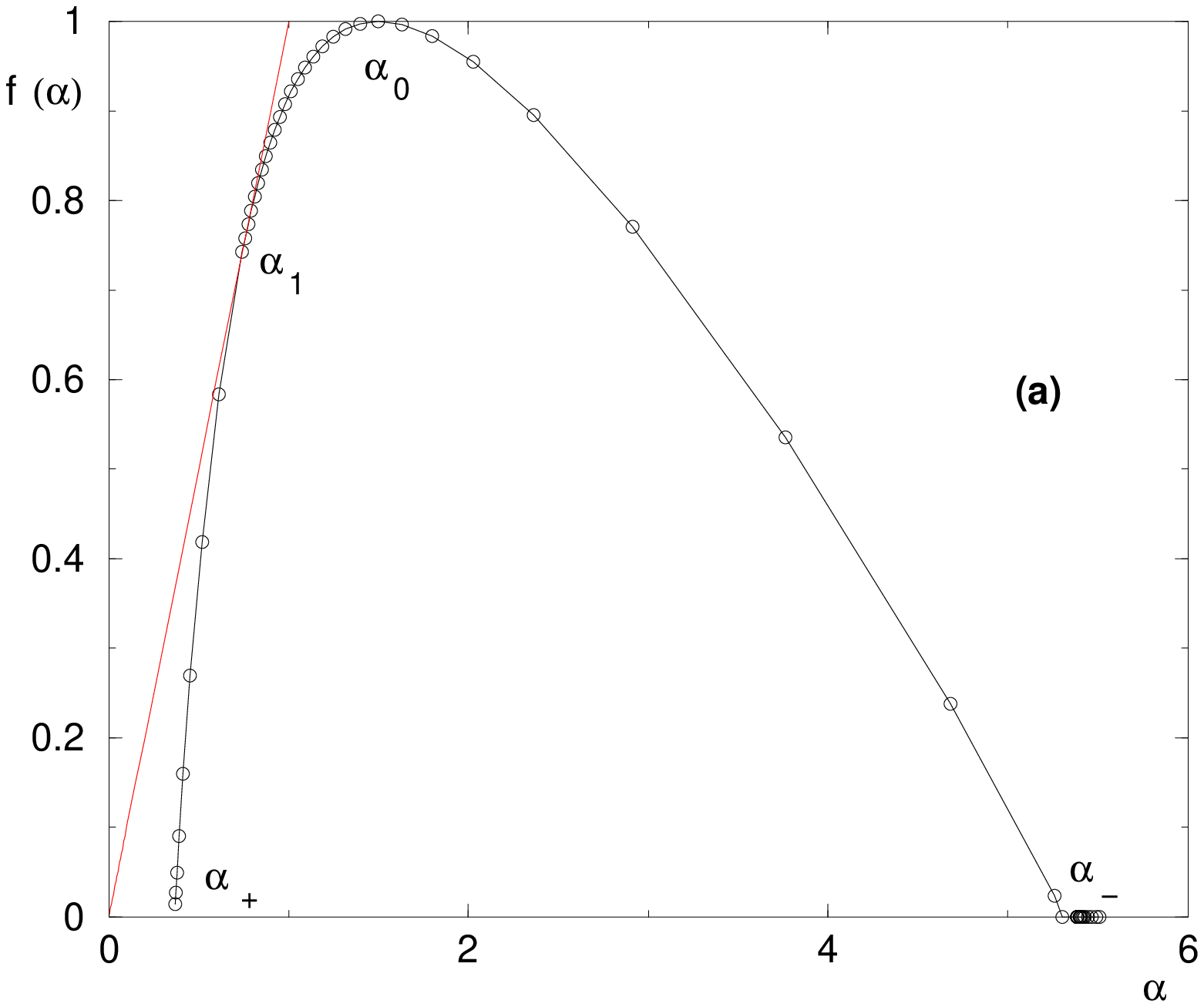}
\hspace{1cm}
 \includegraphics[height=6cm]{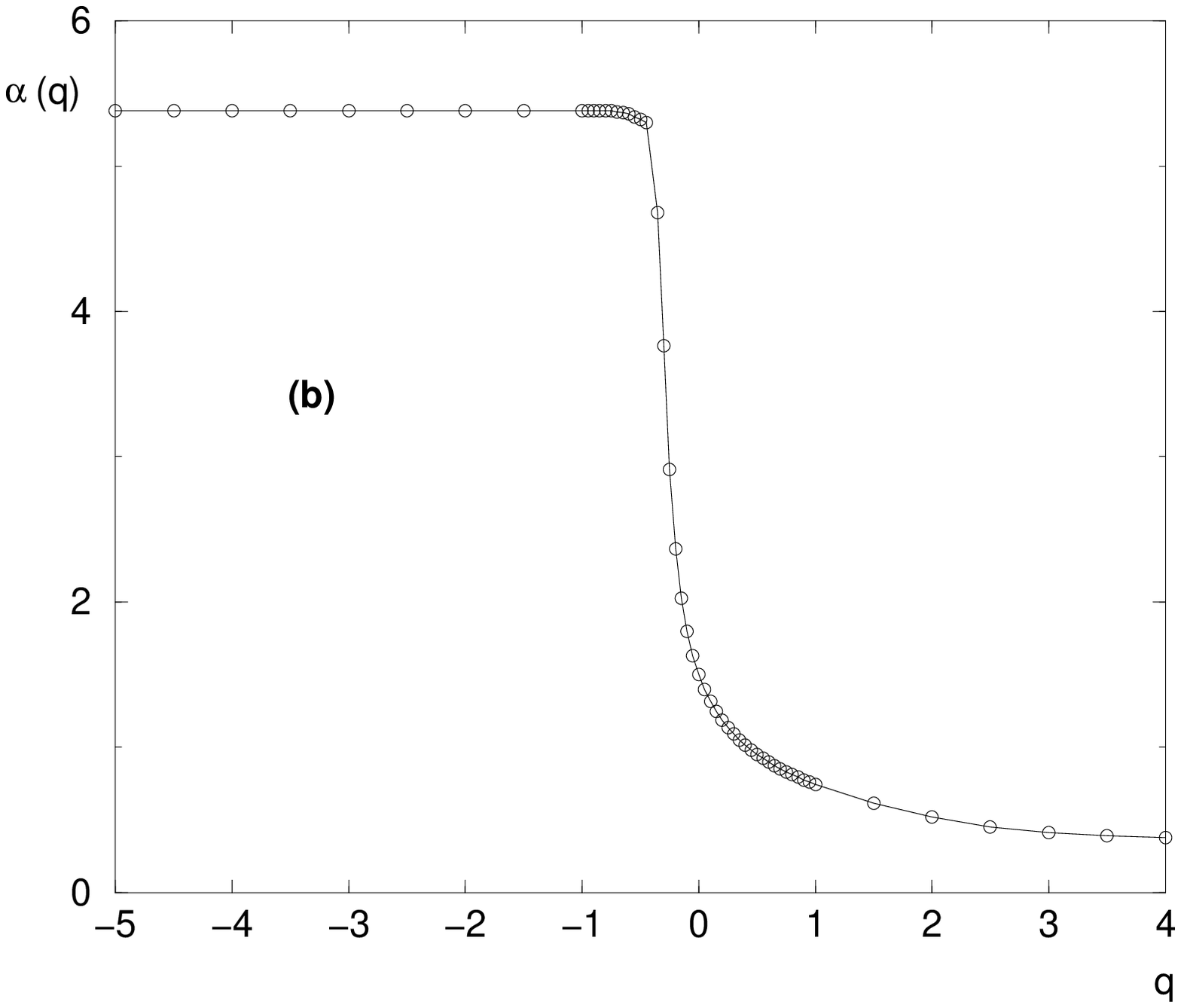}
\caption{ Cauchy disorder $W=0.5$
(a) The singularity spectrum  $f(\alpha)$  has for termination points $\alpha_+ \simeq 0.37$ and $\alpha_- \simeq 5.3$,
for typical value $\alpha_0 \simeq 1.5$ and for tangent point $\alpha_1=f(\alpha_1) \simeq 0.74$
(b) The corresponding $\alpha(q)$ saturates at a value around $q_+ \simeq 3$.
}
\label{figcauchywd=0.5}
\vspace{1cm} 
 \includegraphics[height=6cm]{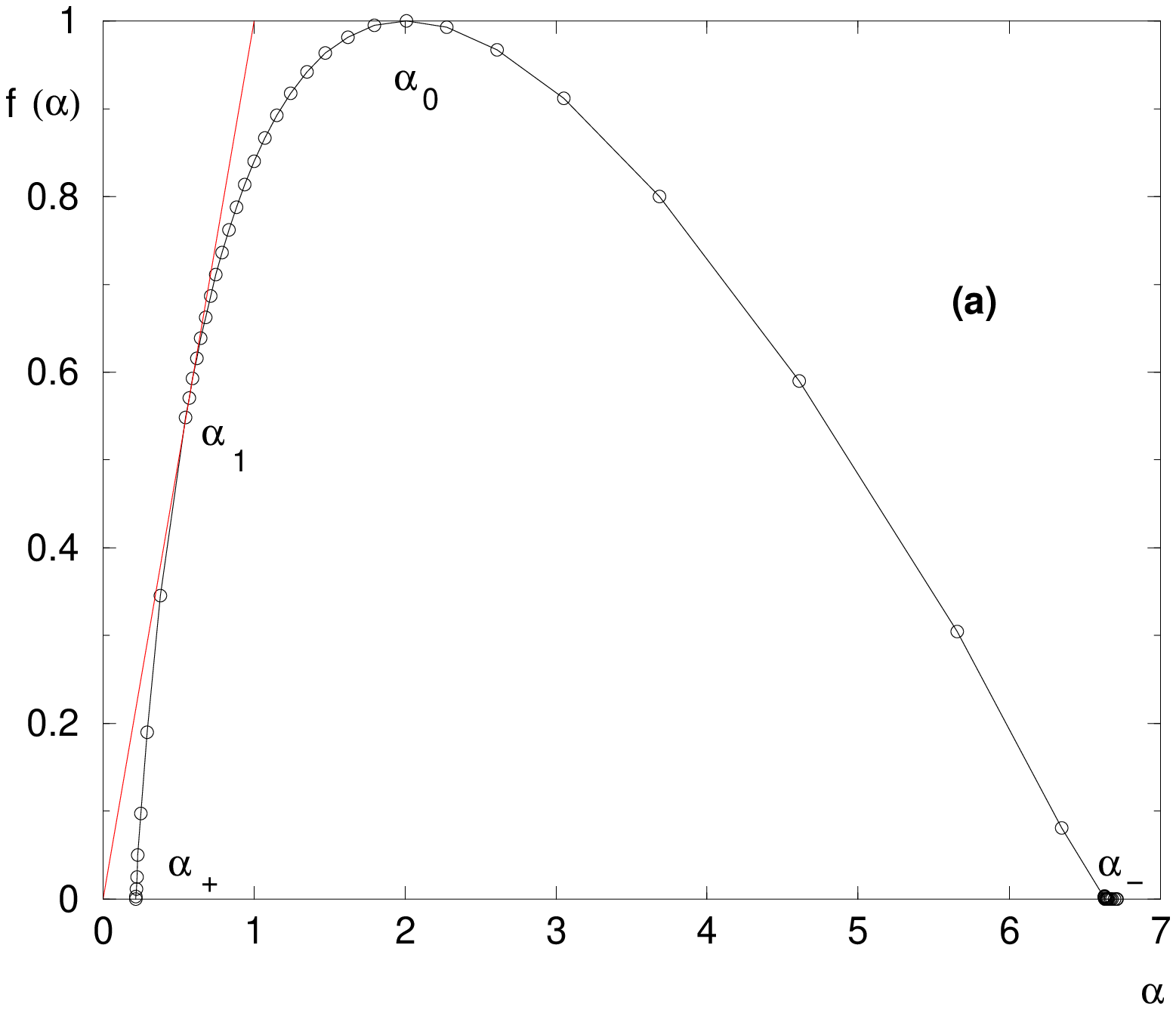}
\hspace{1cm}
 \includegraphics[height=6cm]{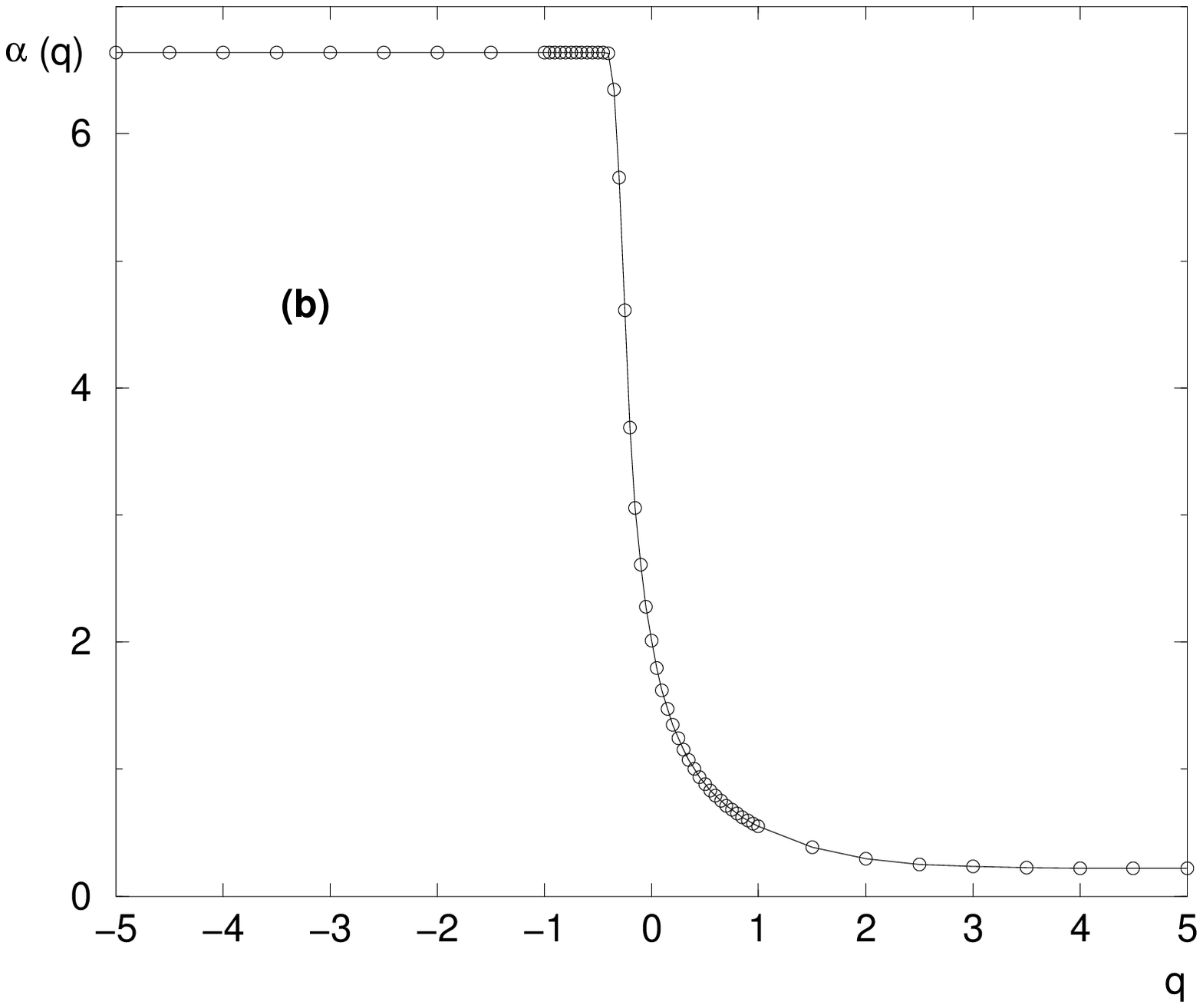}
\caption{ Cauchy disorder $W=1$
(a) The singularity spectrum  $f(\alpha)$  has for termination points $\alpha_+ \simeq 0.21$ and $\alpha_- \simeq 6.63$,
for typical value $\alpha_0 \simeq 2$ and for tangent point $\alpha_1=f(\alpha_1) \simeq 0.55$
(b) The corresponding $\alpha(q)$ saturates at a value around $q_+ \simeq 2$.
  }
\label{figcauchywd=1}
\end{figure}

In the delocalized phase $W<W_c$, we find that the left termination point is strictly positive
$\alpha_+(W)>0$ and is associated with a moment index $q_+(W)>1$.
Two examples of our numerical data are shown on Fig. \ref{figcauchywd=0.5} and \ref{figcauchywd=1}
corresponding to $W=0.5$ and $W=1$ respectively.

\subsection{ Critical point }

\begin{figure}[htbp]
 \includegraphics[height=6cm]{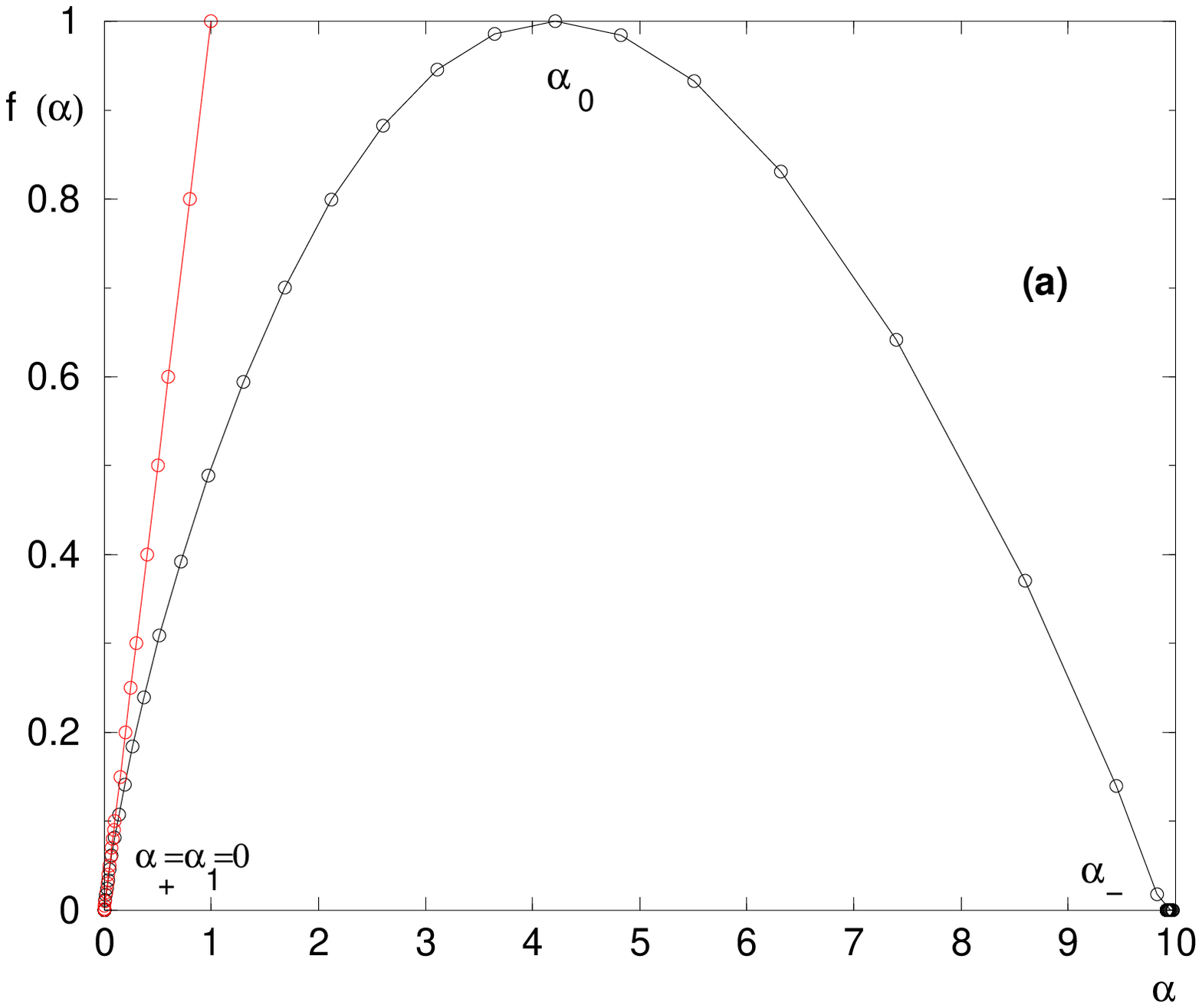}
\hspace{1cm}
 \includegraphics[height=6cm]{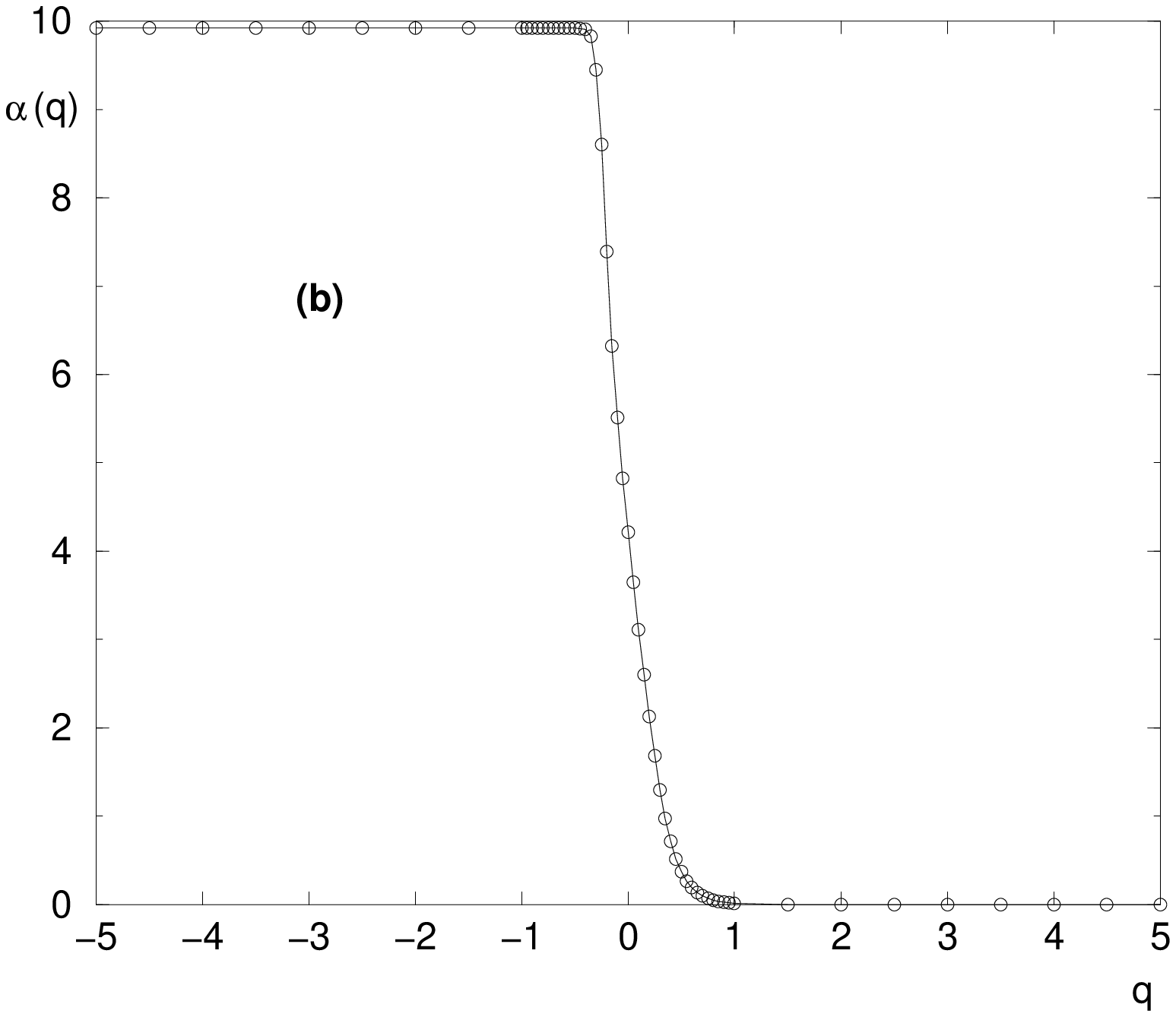}
\caption{ Cauchy disorder $W=4$
(a) The singularity spectrum  $f(\alpha)$ has for termination points $\alpha_+ \simeq 0$ and $\alpha_- \simeq 9.92$,
for typical value $\alpha_0 \simeq 4.21$ and for tangent point $\alpha_1=f(\alpha_1) \simeq 0$
(b) The corresponding  $\alpha(q)$ saturates at the value $q_+ \simeq 1$.
  }
\label{figcauchywd=4}
\end{figure}

At criticality, the left termination point vanishes $\alpha_+(W_c)=0$
together with the tangent point $\alpha_1=f(\alpha_1) = 0$, as shown on Fig. \ref{figcauchywd=4}
corresponding to $W=4$. The corresponding saddle-point $\alpha(q)$ saturates at the value
$q_+(W_c) \simeq 1$.

\subsection{ Localized phase }

\begin{figure}[htbp]
 \includegraphics[height=6cm]{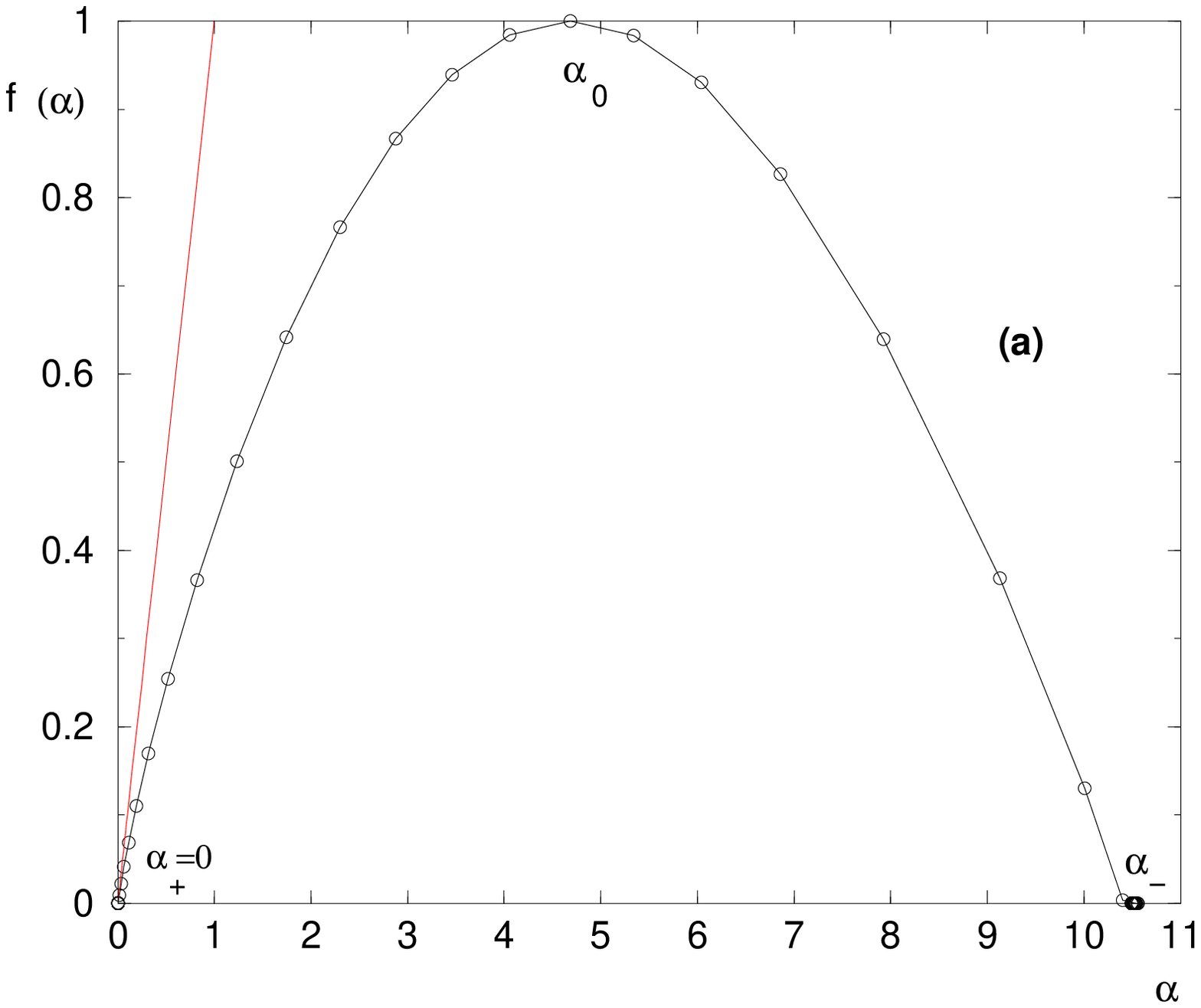}
\hspace{1cm}
 \includegraphics[height=6cm]{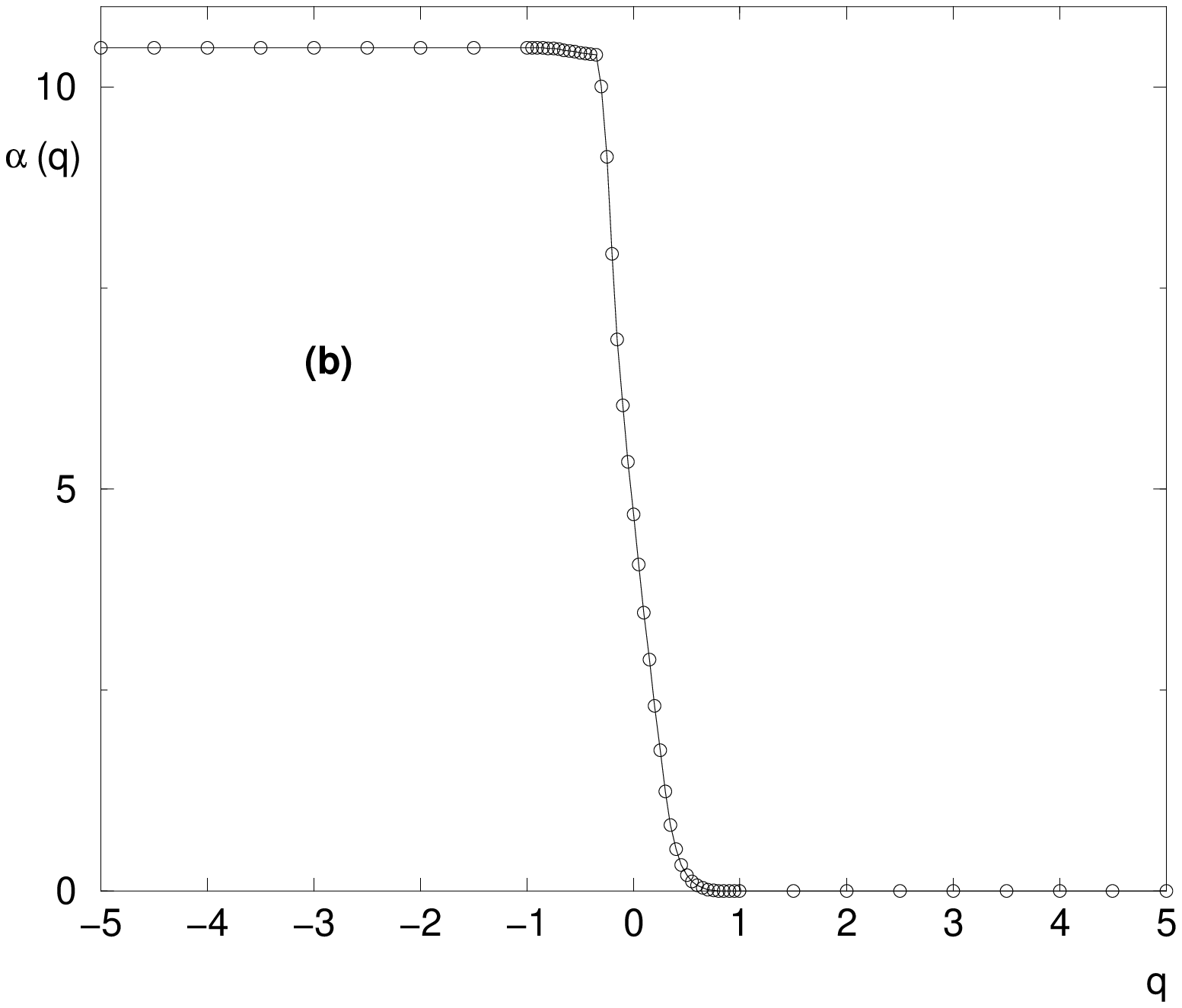}
\caption{ Cauchy disorder $W=6$
(a) The singularity spectrum  $f(\alpha)$  has for termination points $\alpha_+ = 0$ and $\alpha_- \simeq 10.48$,
and for typical value $\alpha_0 \simeq 4.68$
(b) The corresponding  $\alpha(q)$  saturates at the value $q_+ \simeq 0.7$.
  }
\label{figcauchywd=6}
\vspace{1cm} 
 \includegraphics[height=6cm]{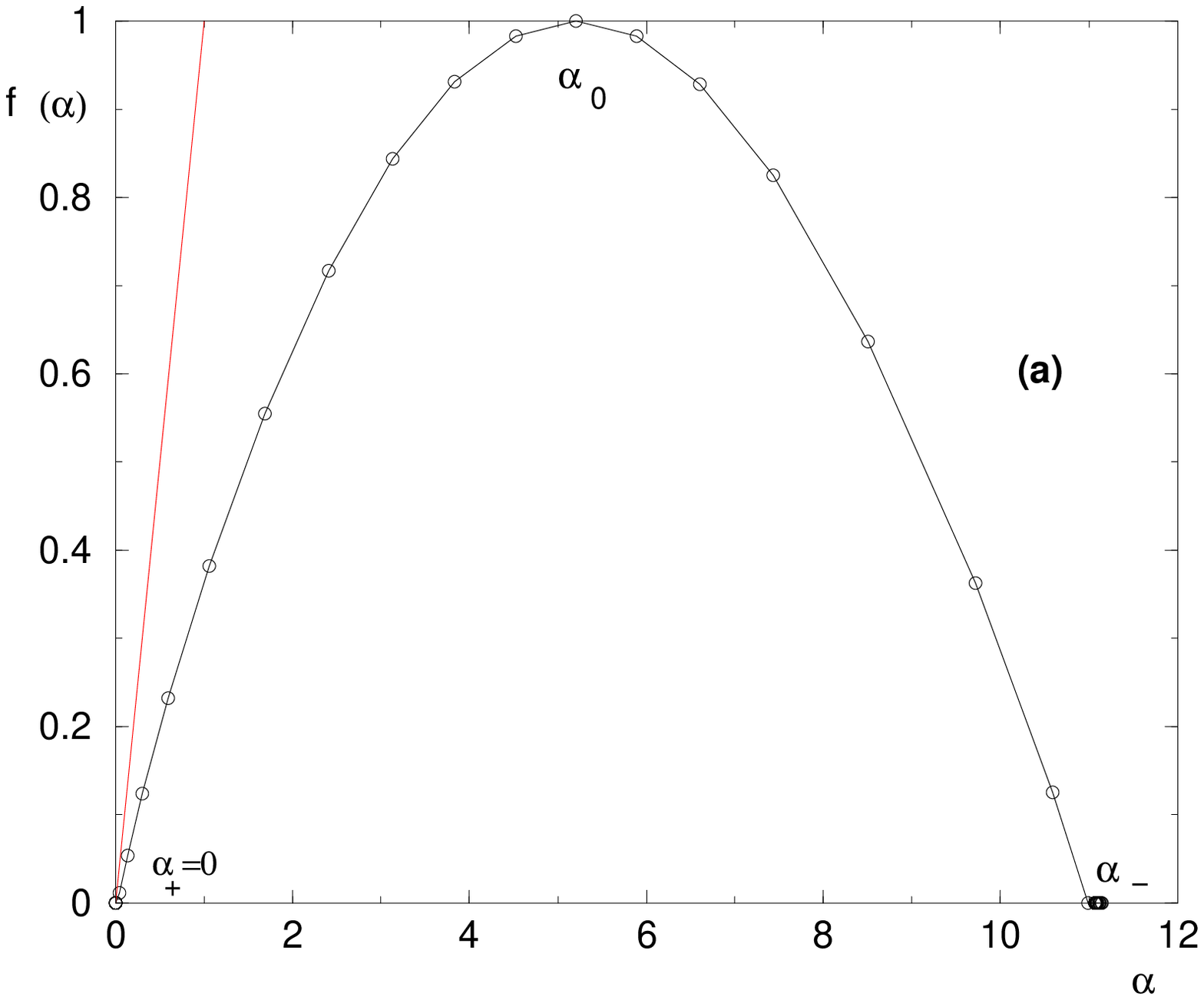}
\hspace{1cm}
 \includegraphics[height=6cm]{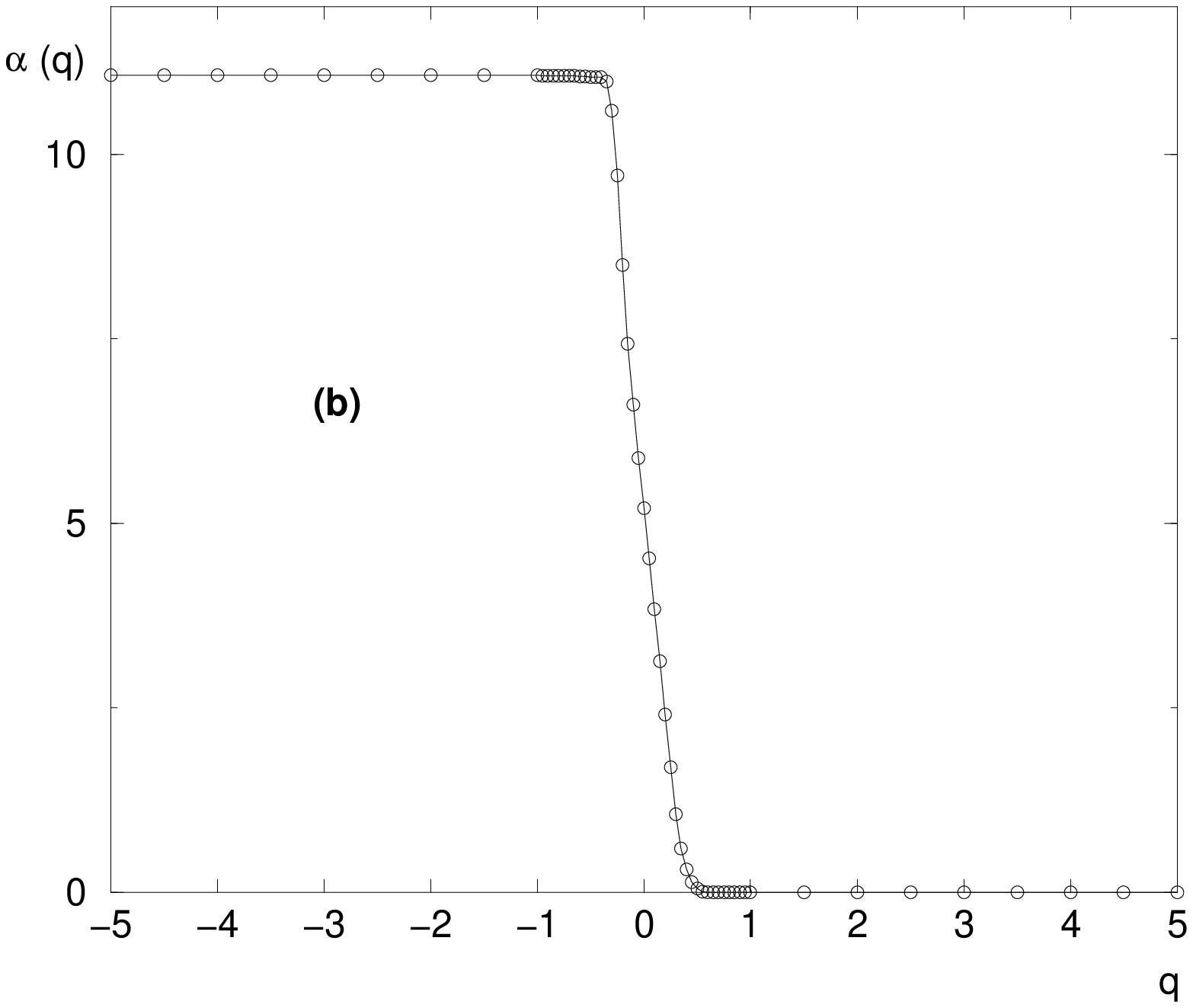}
\caption{ Cauchy disorder $W=10$
(a) The singularity spectrum  $f(\alpha)$  has for termination points $\alpha_+ = 0$ and $\alpha_- \simeq 11.06$,
and for typical value $\alpha_0 \simeq 5.2$
(b) The corresponding  $\alpha(q)$  saturates at the value $q_+ \simeq 0.5$.
  }
\label{figcauchywd=10}
\end{figure}

 In the localized phase $W>W_c$, the 
vanishing left termination point $\alpha_+(W_c)=0$ is associated
with some moment index $q_+(W)<1$, as shown on Fig. \ref{figcauchywd=6} and \ref{figcauchywd=10}
corresponding to $W=6$ and  $W=10$ respectively.

\section{ Conclusions }

\label{sec_conclusion}

In this paper, we have studied the multifractal properties of the Landauer transmission
for the Anderson localization tight-binding model on the Cayley tree within
the Miller-Derrida scattering geometry.
We have explained why, in contrast to finite dimensions where disordered systems display multifractal statistics only at criticality, the tree geometry induces multifractal statistics
for disordered systems also off criticality. As an example, we have recalled in the Appendix the exact results concerning the Directed Polymer on the Cayley tree. 
We have presented numerical results for 
the typical multifractal singularity spectrum $f(\alpha)$
 of the channels  weights as a function of the disorder strength $W$,
both the the Box distribution and the Cauchy distribution of disorder.
Our main conclusion concerns the left-termination point $\alpha_+(W)$.
In the delocalized phase $W<W_c$, $\alpha_+(W)$ is
strictly positive $\alpha_+(W)>0$ and is associated with a moment
index $q_+(W)>1$. At criticality, it vanishes
$\alpha_+(W_c)=0$ and is associated with the moment index
$q_+(W_c)=1$. In the localized phase $W>W_c$, $\alpha_+(W)=0$ is associated with some moment index
$q_+(W)<1$.
These properties of the delocalized and localized phases are thus qualitatively similar 
to the exact results concerning the Directed Polymer on the Cayley tree.


\appendix

\section{ Reminder on the Directed Polymer on the Cayley tree }

\label{sec_DP}

\subsection{Reminder on the thermodynamics} 

The Directed Polymer on a Cayley tree with disorder
has been introduced in \cite{Der_Spo}
 as a mean-field version
of the Directed Polymer in a random medium \cite{Hal_Zha}.
The model is defined by the partition function
\begin{eqnarray}
Z_N= \sum_{{\cal C}} e^{- \beta E({\cal C})}
\label{zcayley} 
\end{eqnarray}
where the $K^N$ configurations $\cal C$ are the paths of $N$ steps
on a Cayley tree with
coordination number $K$. The energy $E({\cal C})$ of a path
is the sum of the energies of the visited bonds.
Each bond has a random energy
drawn independently, for instance with the Gaussian distribution
\begin{eqnarray}
\rho(\epsilon) = \frac{1}{\sqrt{2 \pi}} e^{- \frac{\epsilon^2}{2} }
\label{rhocayley} 
\end{eqnarray}
This model presents many similarities \cite{Der_Spo,Coo_Der}
 with the Random Energy Model, introduced by Derrida
in the context of spin glasses \cite{rem}.
It presents a freezing transition at
\begin{eqnarray}
T_c= \frac{1}{ \sqrt{ 2 \ln K } }
\label{tccayley}
\end{eqnarray}
The free energy per step $\phi(T)$ coincides with the annealed free energy above $T_c$
and is completely frozen below \cite{Der_Spo,Coo_Der}
\begin{eqnarray}
\phi(T) && =\phi_{ann}(T) = - T \ln K - \frac{1}{2 T}
=- \frac{T} { 2 T_c^2} - \frac{1}{2 T}
 {\rm \ \ \ for \ \ }  T \geq
T_c \\
\phi(T) &&  = - \frac{1}{T_c}  {\rm \ \ \ for \ \ }  T \leq T_c
\label{fcayley}
\end{eqnarray}

\subsection{Reminder on the finite weights statistics in the frozen phase }

The configurations weights in the partition function (Eq. \ref{zcayley})
are defined as
\begin{equation}
w_{{\cal C}} = \frac{ e^{- \beta E_{{\cal C}}} }{Z_N(\beta)}
\label{defwi}
\end{equation}
The moments 
\begin{equation}
Y_q(N)=\sum_{i=1}^{K^N} w_{{\cal C}}^q
\label{defyq}
\end{equation}
have finite disorder-averages in the frozen phase $\mu(T)=T/T_c<1 $
for values $q>\mu(T)=T/T_c$ \cite{Der} 
\begin{equation}
\overline{Y_q}= \frac{\Gamma(q-\mu(T))}{\Gamma(q) \Gamma(1-\mu(T))}
\ \ {\rm  with } \ \ \mu(T)=\frac{T}{T_c}
\label{yqrem}
\end{equation}
The density $g(w)$ giving rise to these moments
\begin{equation}
\overline{Y_q} = \int_0^1 dw w^q g(w)
\label{lienyqf}
\end{equation}
reads  \cite{Der}
\begin{equation}
g(w)= \frac{w^{-1-\mu} (1-w)^{\mu-1} }{ \Gamma(\mu) \Gamma(1-\mu)}
\label{densitew}
\end{equation}
and represents the averaged number of terms of weight $w$.
This density is non-integrable as $w \to 0$, because in the limit
 $N \to \infty$, the number of terms of vanishing weights diverges.
The normalization corresponds to 
\begin{equation}
\overline{Y_{q=1}} = \int_0^1 dw w g(w) =1
\end{equation}

\subsection{Reminder on multifractal properties of the weights } 

In the non-frozen phase, the 
$Y_q$ of Eq. \ref{defyq} vanish with the number $M=K^N$ of configurations
as power-laws with non-trivial exponents for averaged and typical values,
and it is convenient to introduce the multifractal
formalism of Eqs \ref{tauq}, \ref{saddle}, \ref{legendre}.
In the frozen phase, the finite asymptotic values obtained for
$q>\mu(T)$ in Eq. \ref{lienyqf}
correspond to $\tau_q=0$, but it is nevertheless interesting
to define the multifractal exponents of Eq. \ref{tauq} 
for $\vert q \vert <\mu$.
The fact that the multifractal formalism is appropriate
to describe the weights statistics 
for all values of $T$, and its exact computation
is explained in detail in \cite{mudry1,mudry2,carpentier_XY,carpentier_log,fyodorov_jpb,fyodorov,fyodorov_pld_ar,fyodorov_lecture}.
Here we simply recall the main results, and we refer to 
 \cite{mudry1,mudry2,carpentier_XY,carpentier_log,fyodorov_jpb,fyodorov,fyodorov_pld_ar,fyodorov_lecture} for more details and discussions.
The main point is that the moments $Y_q$ can be rewritten as in terms of the partition
functions at inverse temperatures $\beta$ and $\vert q \vert \beta$
\begin{eqnarray}
Y_q(M=K^N) \equiv \frac{ \sum_{i=1}^{K^N} e^{- q \beta E_{i}} }{ 
\left( \sum_{i=1}^{K^N} e^{- \beta E_{i} } \right)^q}
= \frac{ Z_N( \vert q \vert \beta) }{(Z_N(\beta))^q} = 
e^{ -  \beta ( \vert q \vert F_N( \vert q \vert\beta) - q F_N(\beta))}
\label{yqzz}
\end{eqnarray}
so that the typical exponents of Eq. \ref{tauq}
are directly given in terms of the free-energy per step of Eq. \ref{fcayley}
\begin{eqnarray}
 \tau^{typ}(q) = \frac{  \beta}{\ln K}
 \left[ \vert q \vert \phi(\vert q \vert \beta) - q \phi(\beta) \right]
\label{tautypfreeenergy}
\end{eqnarray}
In the following, we thus quote the final results for the typical singularity spectrum $f(\alpha)$
in the various phases, and we present the numerical results obtained on trees
of sizes $10 \leq N \leq 22$ to show that such sizes are sufficient to obtain 
reliable results by comparison with the exact forms
(more details on the numerical procedure can be found in Appendix \ref{app_numerics}).

\begin{figure}[htbp]
 \includegraphics[height=5cm]{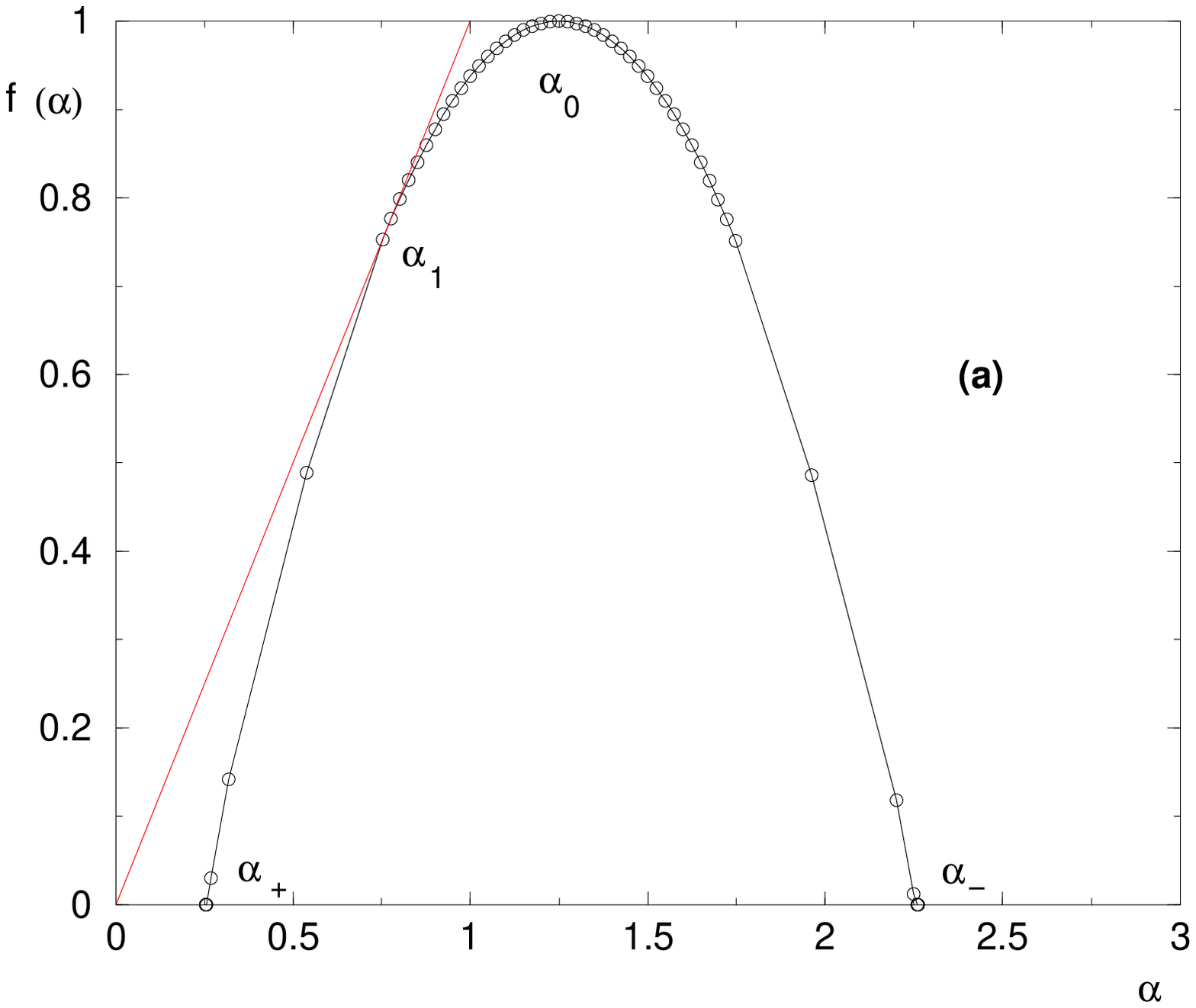}
\hspace{1cm}
 \includegraphics[height=5cm]{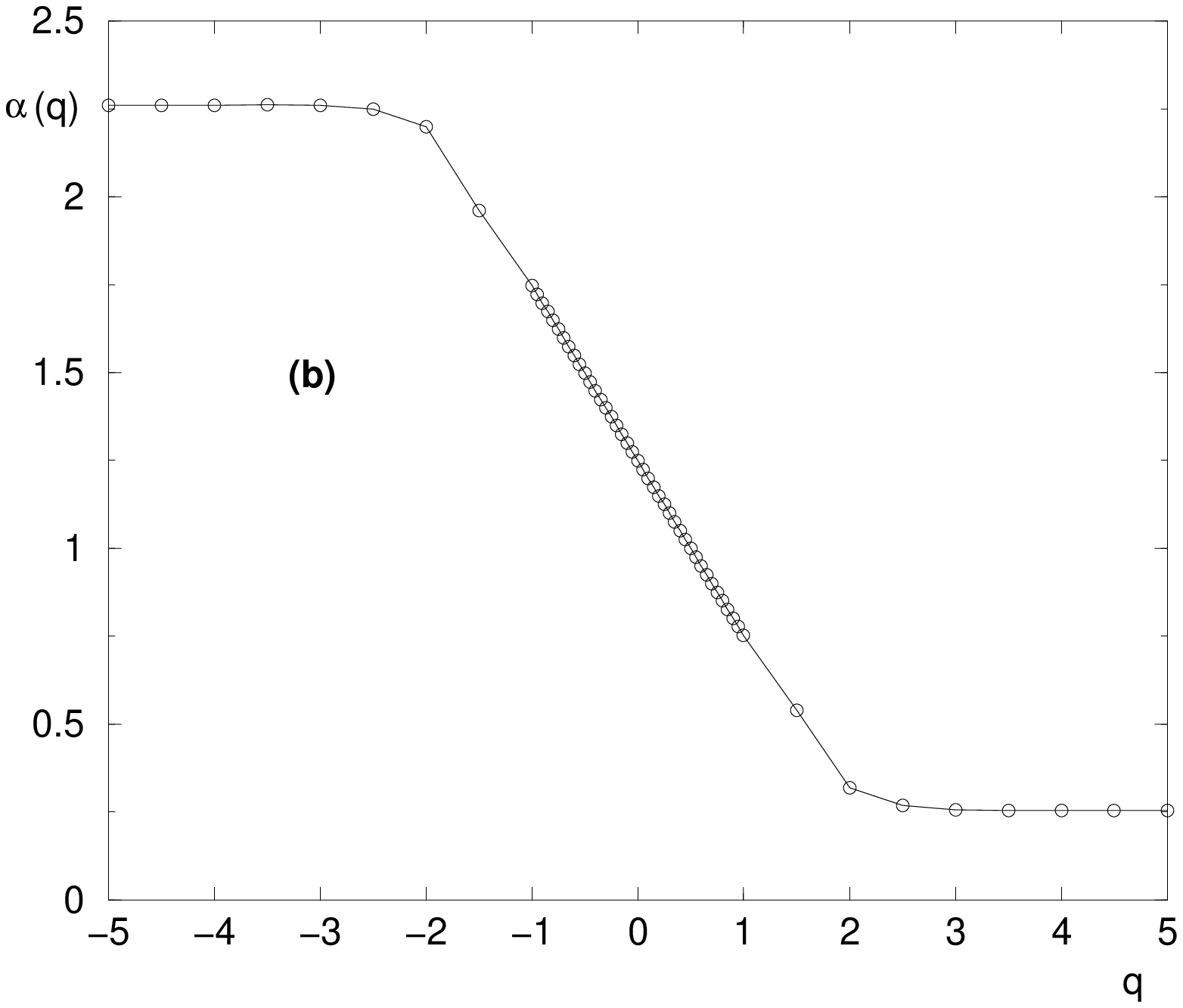}
\caption{ Directed Polymer in the non-frozen phase at $\mu \equiv T/T_c=2$ 
(see section \ref{DPnonfrozen}) :
(a) Singularity spectrum $f(\alpha)$ : the terminating points are $\alpha_+=0.25 $ and $\alpha_-=2.25 $, the typical value is $\alpha_0=1.25$, the line $\alpha=f(\alpha)$
is tangent at $f(\alpha_1)=\alpha_1=0.75$ .
(b) The saddle-point $\alpha(q) $ remains frozen at $\alpha_+$ for $q>q_+=2$ and to
$\alpha_-$ for $q<q_-=-2$. }
\label{figDPmu2}
\vspace{1cm}
\includegraphics[height=5cm]{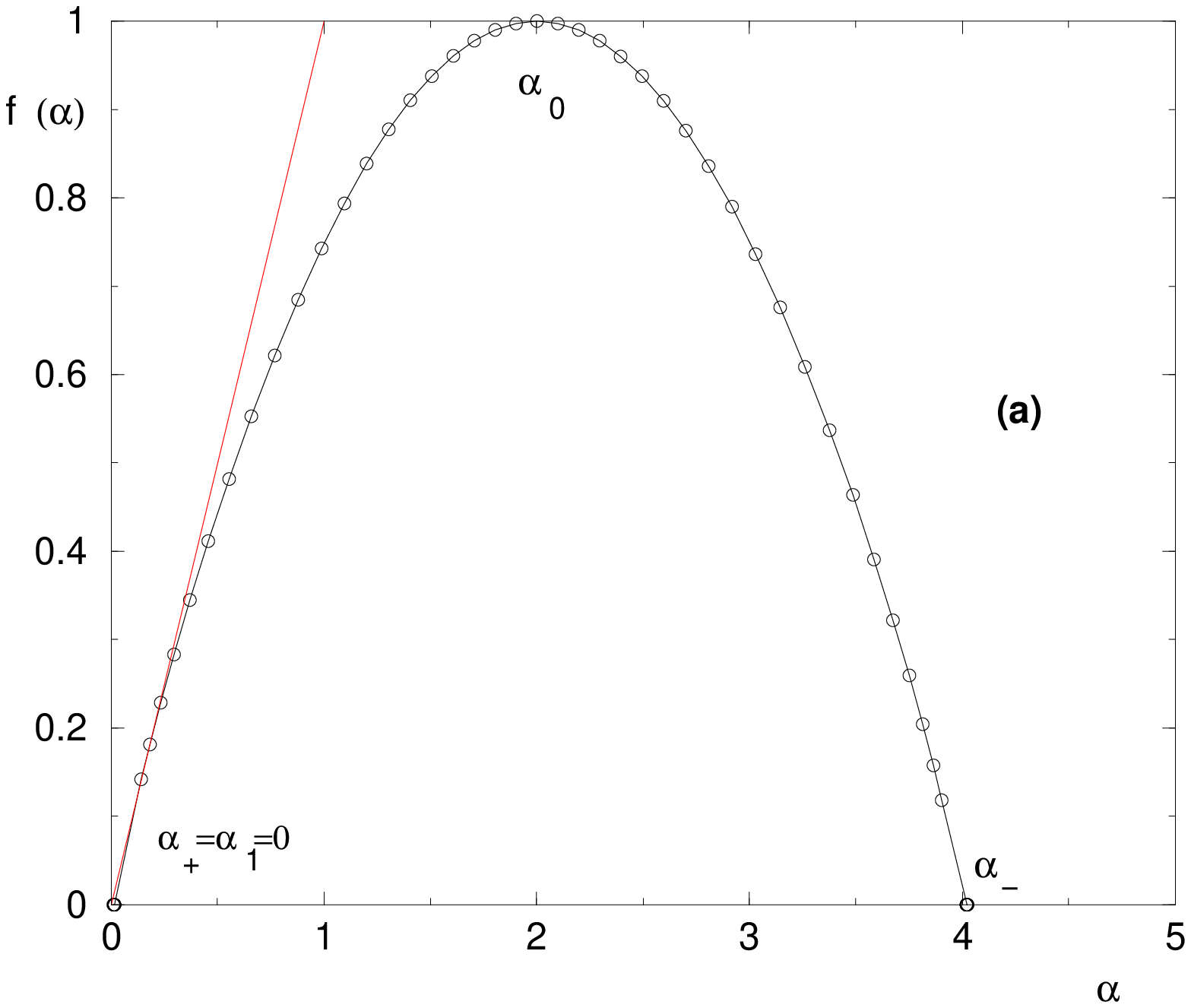}
\hspace{1cm}
 \includegraphics[height=5cm]{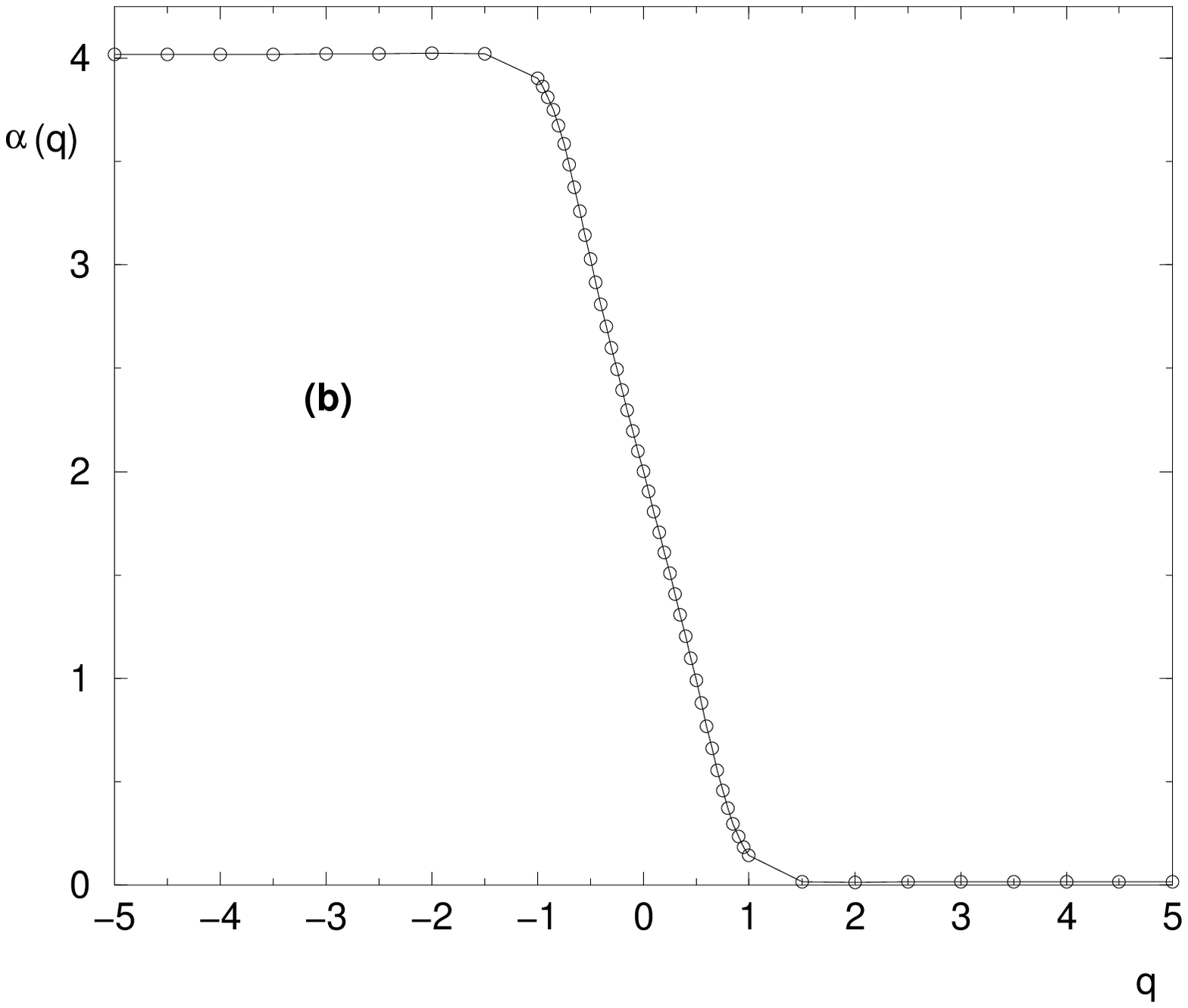}
\caption{ Directed Polymer at the critical point for $\mu \equiv T/T_c =1$
 (see section \ref{DPcriti} ) :
 (a) Singularity spectrum $f(\alpha)$ : the terminating points are $\alpha_+=0$ and $\alpha_-=4 $, the typical value is $\alpha_0=2$, the line $\alpha=f(\alpha)$
is tangent at the origin $f(\alpha_1)=\alpha_1=0$ .
(b) The saddle-point $\alpha(q) $ remains frozen at $\alpha_+$ for $q>q_+=1$ and to
$\alpha_-$ for $q<q_-=-1$.}
\label{figDPmu1}
\vspace{1cm}
\includegraphics[height=5cm]{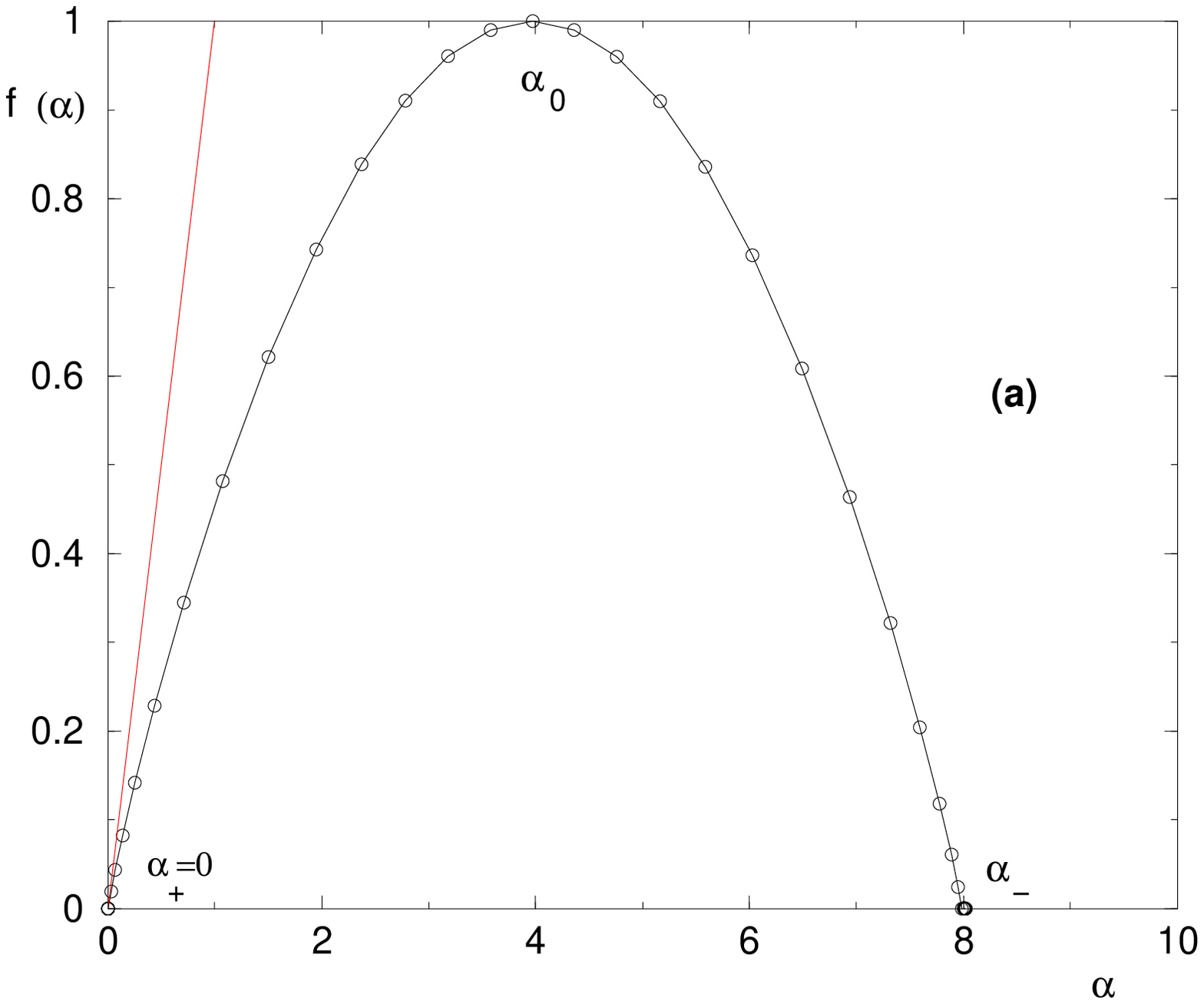}
\hspace{1cm}
 \includegraphics[height=5cm]{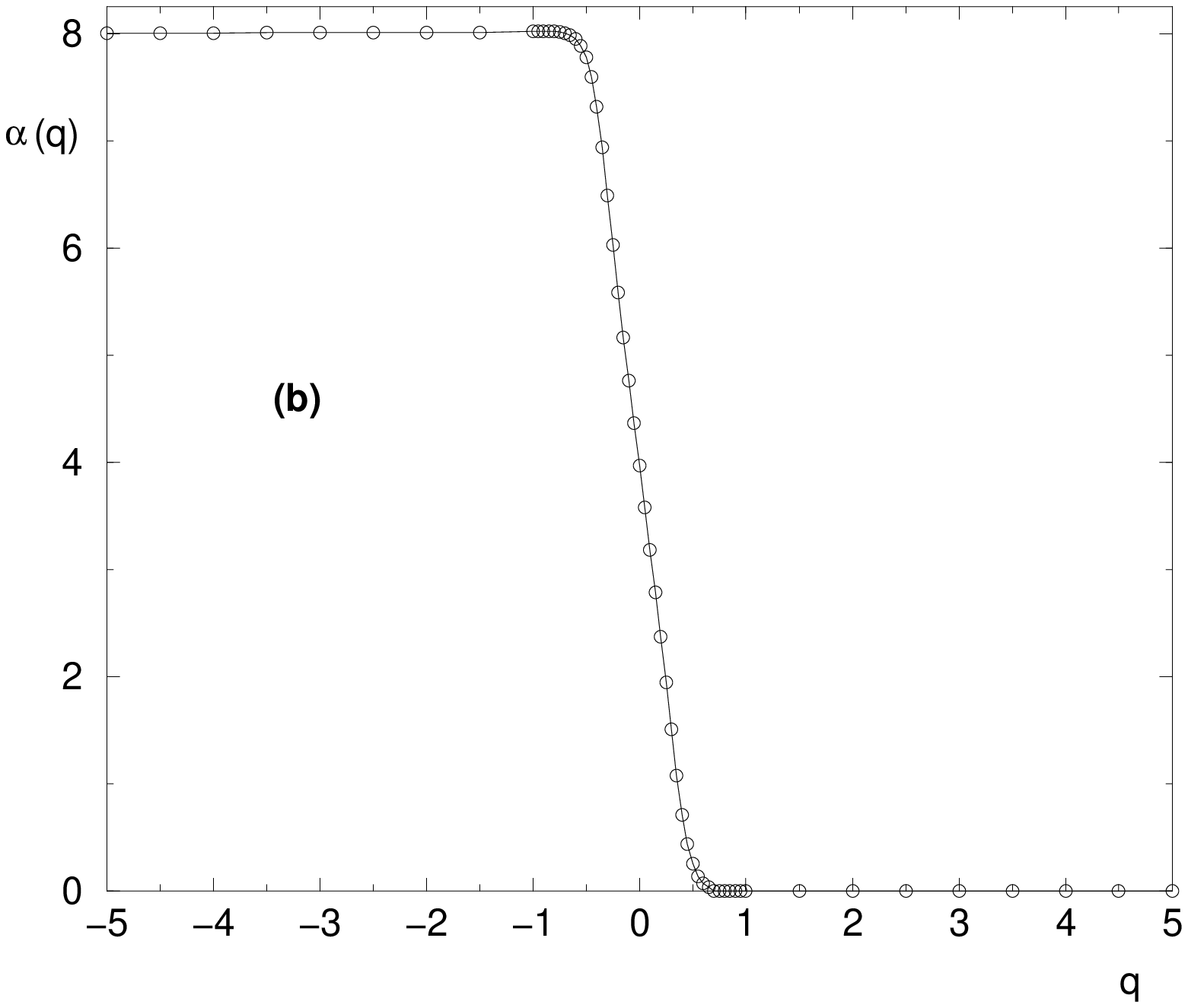}
\caption{ Directed Polymer in the frozen phase at $\mu \equiv T/T_c=0.5$ (see section \ref{DPfrozen} ) : 
(a) Singularity spectrum $f(\alpha)$ : the terminating points are $\alpha_+=0 $ and $\alpha_-=8 $, the typical value is $\alpha_0=4$.
(b) The saddle-point $\alpha(q) $ remains frozen at $\alpha_+$ for $q>q_+=0.5$ and to
$\alpha_-$ for $q<q_-=-0.5$. }
\label{figDPmu0.5}
\end{figure}

\subsubsection{Non-Frozen Phase $\mu \equiv \frac{T}{T_c} >1$ }

\label{DPnonfrozen}

In the non-frozen phase, the left and right termination points read 
\begin{eqnarray}
\alpha_+  && =  \left(1  - \frac{1  }{ \mu }  \right)^2  \nonumber \\
\alpha_- && =  \left(1  + \frac{1  }{ \mu }  \right)^2  
\label{nonfrozenalphamoins}
\end{eqnarray}
and 
 the typical singularity spectrum is exactly Gaussian on the interval
$\alpha_+<\alpha < \alpha_-$ where it exists 
\begin{eqnarray}
f^{typ}_{T>T_c}(\alpha) =  \frac{\mu^2}{4} (\alpha-\alpha_+ ) (\alpha_- - \alpha )
\label{nonfrozenfalphagauss}
\end{eqnarray}

The terminating values $\alpha_{\pm}$ are associated with the values  $q_{\pm}=\pm \mu$.
In the interval $q_-=-\mu \leq q \leq q_+=+\mu$, the value $\alpha(q)$ dominating the saddle point calculation
of Eq. \ref{legendre} is simply linear in $q$
\begin{eqnarray}
\alpha(q_-=-\mu \leq q \leq q_+=+\mu) =  \left(1+\frac{1}{\mu^2}  \right) - \frac{2 q}{ \mu^2}
\label{nonfrozenfalphasaddle}
\end{eqnarray}
The typical value corresponding to $q=0$ in Eq. \ref{nonfrozenfalphasaddle}
\begin{eqnarray}
\alpha_0  =  \left(1+\frac{1}{\mu^2}  \right)
\label{nonfrozenalphatyp}
\end{eqnarray}
is the point where the singularity spectrum reaches its maximum $f(\alpha_0)=1$
(Eq. \ref{analphaq0}).
Finally the value $q=1$ where the singularity spectrum is tangent to the line $\alpha=f(\alpha)$ (Eq. \ref{tanalphaq1}) correspond to (Eq. \ref{nonfrozenfalphasaddle})
\begin{eqnarray}
\alpha_1 = 1-\frac{1}{\mu^2} = f(\alpha_1)
\label{tanalphaq1dpnonfrozen}
\end{eqnarray}

The numerical results shown on Fig. \ref{figDPmu2} are in agreement
 with these expressions for $\mu=2$. 

 \subsubsection{ Critical point $\mu \equiv \frac{T}{T_c} =1$ }

\label{DPcriti}

In the limit $\mu \equiv \frac{T}{T_c} \to 1^+$, the above singularity spectrum
has the following properties : 
the left terminating point $\alpha_+$ of Eq. \ref{nonfrozenalphamoins} vanishes
\begin{eqnarray}
\alpha_+  =0
\label{critialphaplus}
\end{eqnarray}
together with the tangent point to the line $\alpha=f(\alpha)$
 (Eq. \ref{tanalphaq1dpnonfrozen})
\begin{eqnarray}
\alpha_1  = f(\alpha_1) =0
\label{tanalphaq1dpcriti}
\end{eqnarray}
The corresponding numerical results for $\mu=1$ are shown on Fig. \ref{figDPmu1}. 

\subsubsection {Frozen Phase $\mu \equiv \frac{T}{T_c} <1$ }

\label{DPfrozen}

In the frozen phase, the left termination point is zero
\begin{eqnarray}
\alpha_+   = 0
\label{frozenalphaplus}
\end{eqnarray}
and the right termination point is
\begin{eqnarray}
\alpha_-  =  \frac{4}{\mu} 
\label{frozenalphamoins}
\end{eqnarray}
On the interval $\alpha_+=0<\alpha < \alpha_-$ where it exists,
the typical singularity spectrum is again exactly Gaussian 
\begin{eqnarray}
f^{typ}_{T<T_c}(\alpha) = \frac{\mu^2}{4} \alpha (\alpha_- - \alpha )
\label{frozenfalphagauss}
\end{eqnarray}

The left terminating value $\alpha_+=0$ is reached for $q_+=\mu$
and the right terminating value $\alpha_-$ is reached for $q_-=-\mu$.
 In the interval $q_-=-\mu \leq q \leq q_+=+\mu$, the value $\alpha(q)$ dominating the saddle point calculation
of Eq. \ref{legendre} is again  linear in $q$
\begin{eqnarray}
\alpha(q_-=-\mu \leq q \leq q_+=+\mu) = \frac{2}{\mu}  - \frac{2 q}{ \mu^2}
\label{frozenfalphasaddle}
\end{eqnarray}
The typical value corresponding to $q=0$ in Eq. \ref{frozenfalphasaddle} is
\begin{eqnarray}
 \alpha_0 =  \frac{2}{\mu} 
\label{frozenalphatyp}
\end{eqnarray}
Note that here the value $q=1$ is already is the frozen region $1>q_+=\mu$,
so that the singularity spectrum is completely below the line $\alpha=f(\alpha)$.
The numerical results shown on Fig. \ref{figDPmu0.5} are in agreement
 with these expressions for $\mu=0.5$. 

\section{ Details concerning the numerical evaluation of $f(\alpha)$ }

\label{app_numerics}

In each sample with $N$ generations, 
we compute the $M=K^N$ weights of Eq. \ref{wjtrans}, from which we obtain immediately
the parameters $I_q$ of Eq. \ref{iqdef}. The corresponding typical exponents
$\tau^{typ}(q)$ of Eq. \ref{tauq} are then obtained by the following 
 three parameters fit of the average over disordered samples 
\begin{eqnarray}
\overline{ \ln I_q(M) } \opsimeq  -\tau_q ^{typ} \ln M + c_0 \ln (\ln M) +c_0' 
\label{FitIq}
\end{eqnarray}
i.e. $\tau_q^{typ}$ is obtained as the coefficient of the leading linear term.
The presence of the subleading term $(c_0)$ is important 
only in the regions where $\tau_q$ nearly vanishes $\tau_q \sim 0$, 
whereas in the regions where $\tau_q$ is not small,
a direct linear fit could be acceptable and would give nearly the same numerical value
for $\tau_q$.

The typical multifractal spectrum $f^{typ}(\alpha)$ could in principle be obtained
from $\tau^{typ}(q) $ by some numerical procedure to perform the
Legendre transform of Eq. \ref{legendre}, 
but this method has a lot of numerical drawbacks \cite{chh}. We have thus followed 
the standard method of Ref \cite{chh}, with the simplification that we consider only boxes of size $L=1$ in the notation of Ref \cite{chh}, i.e. we do not use any coarse-graining with various box sizes, but we analyse instead the scaling with respect to the total number of boxes $M=K^N$ from our data obtained of various sizes $N$.
The main idea of Ref \cite{chh} is to construct the following normalized
$q-$measures from the initial measure defined by the weights of Eq. \ref{wjtrans}
\begin{eqnarray}
w_j^{(q)} \equiv \frac{ \left[ w_j \right]^q }{ \sum_{j'} \left[ w_{j'} \right]^q }
\label{wjq}
\end{eqnarray}
Of course $q=1$ correspond to the initial measure $w_j^{(q=1)}=w_j $.
The denominator corresponds to $I_q$ of Eq. \ref{iqdef}.
It is then useful to introduce 
\begin{eqnarray}
F_q(M) = -  \sum_{j=1}^{M} w_j^{(q)} \ln w_j^{(q)}  \nonumber \\
A_q(M) = -  \sum_{j=1}^{M} w_j^{(q)} \ln w_j   
\label{FA}
\end{eqnarray}
$F_q(M) $ represents the Shannon entropy of the $q-$ measure, whereas $A_q(M)$ represents the averaged log of the initial weight $\ln w_j$ 
with respect to the $q-$measure. 
From Eq. \ref{wjq}, one obtains immediately the simple relation
\begin{eqnarray}
F_q(M) = q A_q(M) + \ln I_q(M)
\label{FAlegendre}
\end{eqnarray}
which, after the division by the scaling factor $(\ln M)$,
exactly corresponds to the Legendre transform relation of Eq. \ref{legendre}.
Numerically, one only has to compute $I_q$ and $A_q$,
whereas $F_q$ can be immediately obtained from them by Eq. \ref{FA}.
The averages over the disordered samples of the two 
 observables of Eq. \ref{FA} can be analyzed by the following three
parameters fits
\begin{eqnarray}
\overline{F_q(M)} \opsimeq  f_q \ln M + c_1 \ln (\ln M) +c_1' \nonumber \\
\overline{A_q(M)} \opsimeq  \alpha_q \ln M + c_2 \ln (\ln M) +c_2'
\label{FAfits}
\end{eqnarray}
The leading coefficients $(\alpha_q,f_q)$ then constitutes a parametric representation
of the typical singularity spectrum $f(\alpha)$ as $q$ varies.
The presence of the subleading terms $(c_1,c_2)$ are important 
only in the regions where $f_q$ nearly vanishes $f_q \sim 0$, 
whereas in the regions where $f_q$ is not small,
direct linear fits could be acceptable and would give nearly the same numerical values
for $(\alpha_q,f_q)$.

All the singularity spectra $f(\alpha)$ shown on the figures
correspond to the parametric representation $(\alpha_q,f_q)$ obtained by the procedure
just described. We have also presented our corresponding data for $\alpha_q$
to show the freezing phenomena in the parameter $q$.
In Appendix A, we find that the singularity spectra obtained via this numerical analysis are in agreement with the available exact results.

\end{document}